\documentclass[preprint,showpacs,titlepage,aps,prd,tightenlines,amsmath,byrevtex,nofootinbib]{revtex4}
\usepackage{graphicx}

\newcommand{\eV}{\text{eV}}
\newcommand{\km}{\text{km}}

\begin{document}

\preprint{KEK-TH-978}
\preprint{hep-ph/0409109}

\title{
Systematic limits on $\sin^2{2\theta_{13}}$
in neutrino oscillation experiments with multi-reactors
}

\author{H.~Sugiyama}
\email{E-mail: hiroaki_at_post.kek.jp}
\affiliation{Theory Group, KEK, Tsukuba, Ibaraki 305-0801, Japan}

\author{O.~Yasuda}
\email{E-mail: yasuda_at_phys.metro-u.ac.jp}
\affiliation{Department of Physics, Tokyo Metropolitan University,
Hachioji, Tokyo 192-0397, Japan}

\author{F.~Suekane}
\email{E-mail: suekane_at_awa.tohoku.ac.jp}
\affiliation{Research Center for Neutrino Science, Tohoku University,
Sendai, Miyagi 980-8578, Japan}

\author{G.A.~Horton-Smith}
\email{E-mail: gahs_at_phys.ksu.edu}
\affiliation{Kansas State University, Manhattan, KS 66506, USA}

%\date{\today}

%%%%%%%%%%%%%%%%%%%%%%%%%%%%%%%%%%%%%%%%%%%%%%%%%%%%%%%%%%%%%%%%%
%    Abstract
%%%%%%%%%%%%%%%%%%%%%%%%%%%%%%%%%%%%%%%%%%%%%%%%%%%%%%%%%%%%%%%%%
%\hfuzz=25pt
\begin{abstract}
 Sensitivities to $\sin^2{2\theta_{13}}$
without statistical errors (``systematic limit'')
are investigated in neutrino oscillation experiments
with multiple reactors.
Using an analytical approach, we show
that the systematic limit on $\sin^2{2\theta_{13}}$
is dominated by the uncorrelated systematic error
$\sigma_{\text{u}}$
of the detector.
Even in an experiment with multi-detectors and multi-reactors,
it turns out that
most of the systematic errors including the one
due to the nature of multiple sources
is canceled as in the case with a single reactor
plus two detectors,
if the near detectors are placed suitably.
The case of the KASKA plan (7 reactors and 3 detectors) is
investigated in detail, and it is explicitly shown that it
does not suffer from the extra uncertainty
due to multiple reactors.
\end{abstract}

\pacs{14.60.Pq,25.30.Pt,28.41.-i}

\maketitle

%%%%%%%%%%%%%%%%%%%%%%%%

\section{Introduction}

Recently the possibility to measure $\theta_{13}$
by a reactor experiment has attracted much attention%
~\cite{kr2det,Minakata:2002jv,Huber:2003pm,kaska,
Ardellier:2004ui,Shaevitz:2003ws,USreactor,Daya,Anderson:2004pk}.
 To achieve
sensitivity $\sin^2{2\theta_{13}}\sim 0.01$, reduction
of the systematic errors is crucial, and near and far detectors seem
to be necessary for that purpose.  On the other hand, it appears to be
advantageous to do an experiment at a multi-reactor site to gain
statistics and high signal to noise ratio,
and in fact in the most of cases considered in
\cite{kr2det,Minakata:2002jv,Huber:2003pm,kaska,
Ardellier:2004ui,Shaevitz:2003ws,USreactor,Daya,Anderson:2004pk}
there are more than one reactor.
In this paper we discuss the systematic errors
in reactor neutrino oscillation experiments
with multi reactors and multi detectors in an analytical way.
In Sect.2 we discuss the cases with
a single reactor to illustrate our analytical method.
In Sect.3 we consider the cases with $n_r$ reactors
and show that the larger $n_r$ gives totally the smaller contribution
to the sensitivity from the uncorrelated errors
of the fluxes.
 Irrespective of the number of reactors,
if there are more than one detectors,
we can cancel the correlated errors which includes the error of the fluxes.
We emphasize in this paper that
the sensitivity on $\sin^2{2\theta_{13}}$ with vanishing statistical errors
(we refer to the sensitivity as the systematic limit)
is dominated by the uncorrelated
error of the detectors in most cases.
 It is also emphasized that a lot of caution has to be exercised
to estimate the uncorrelated error.
In the appendix we give some details
on how to derive the analytic results used in the main text,
using the equivalence of the pull method and the
covariance matrix approach~\cite{Stump:2001gu,
Botje:2001fx,Pumplin:2002vw,Fogli:2002pt}.
Throughout this paper we do not use the binning of the numbers of events
because the discussions on the uncorrelated bin-to-bin systematic errors
are complicated.  Also we will discuss
only the systematic errors, i.e., we will consider
the case where the statistical errors are negligibly small.

\section{Systematic errors}
To discuss the systematic limit on $\sin^2{2\theta_{13}}$
in neutrino oscillation experiments with multiple reactors,
we have to introduce
the systematic errors of the detectors and
the reactors (fluxes).  There are two kinds of systematic errors of the
detectors, namely, the correlated error $\sigma_c$
and uncorrelated error $\sigma_u$.
The former includes the theoretical uncertainty in the cross section
of detection, etc.\ while the latter does the uncertainty
in the baseline lengths, a portion of measuring the detector volume,
a part of the detection efficiency, etc.
As for the systematic errors of the reactors (fluxes),
the correlated error $\sigma^{(r)}_c$ consists of 
the uncertainties in the spectrum of the $\overline{\nu}_e$ flux, etc.\
whereas the uncorrelated error $\sigma^{(r)}_u$ consists of
the uncertainties in the composition of the fuel, etc.
In this paper we adopt the reference values 
for $\sigma_{\text{c}}$ and $\sigma_{\text{u}}$ used
in \cite{Minakata:2002jv}, where basically the same
reference values as in the Bugey experiment~\cite{Declais:1994su} were assumed.
$\sigma_{\text{c}}$ and $\sigma_{\text{u}}$ can be estimated to be
\begin{eqnarray}
\sigma_{\text{u}}&=&0.8\%/\sqrt{2}=0.6\%\nonumber\\
\sigma_{\text{c}}&=&
\sqrt{(2.7\%)^2-(2.1\%)^2-(0.8\%/\sqrt{2})^2}=1.6\%,
\label{error1}
\end{eqnarray}
where the factor $\sqrt{2}$ appears because the
relative normalization $\sigma_{\text{rel}}$=0.8\%
in \cite{Declais:1994su} is related to $\sigma_{\text{u}}$ by
$\sigma_{\text{rel}}=\sqrt{2}\sigma_{\text{u}}$.
 In the estimation of $\sigma_{\text{c}}$,
we used 2.7\% total error and 2.1\% error of the flux
which are the values in the CHOOZ experiment.
 As for the correlated and uncorrelated errors
of the the flux from the reactors,
we adopt the same reference values as those used by
the KamLAND experiment~\cite{Eguchi:2002dm}
\begin{eqnarray}
\sigma_{\text{c}}^{\text{(r)}}&=&2.5\%,\nonumber\\
\sigma_{\text{u}}^{\text{(r)}}&=&2.3\%.
\label{error2}
\end{eqnarray}
 Note that the word ``correlated'' means just the type of the error,
and then correlated errors exist even if there is no partner.

\section{One reactor\label{onereactor}}
To explain our analytical approach, let us start with the 
simplest case, namely the case with one reactor.

\subsection{One detector}
Let $m$ be the measured number of events at the
detector, $t$ be the theoretical predictions (hypothesis)
to be tested.
$\chi^2$ is defined as
\begin{eqnarray}
\hspace*{-20mm}
\displaystyle
\chi^2&=&\min_{\alpha's}\left\{
\left[{m-t(1+\alpha_{\text{c}}+\alpha_{\text{c}}^{\text{(r)}}
+\alpha_{\text{u}}^{(r)})
\over t\sigma_{\text{u}}}\right]^2
%\right.\nonumber\\
%&{\ }&\left.
+\left({\alpha_{\text{c}} \over \sigma_{\text{c}}}\right)^2
+\left({\alpha_{\text{c}}^{\text{(r)}} \over 
\sigma_{\text{c}}^{\text{(r)}}}\right)^2
+\left({\alpha_{\text{u}}^{\text{(r)}} \over 
\sigma_{\text{u}}^{\text{(r)}}}\right)^2\right\}\nonumber\\
&=&{\left(\displaystyle{m \over t}-1\right)^2 \over
\sigma_{\text{u}}^2+\sigma_{\text{c}}^2
+(\sigma_{\text{c}}^{\text{(r)}})^2
+(\sigma_{\text{u}}^{\text{(r)}})^2},
\label{chi0}
\end{eqnarray}
where $\alpha_u$, $\alpha_c$, $\alpha^{(r)}_c$ and
$\alpha^{(r)}_u$ are the variables of noises to introduce
the systematic errors $\sigma_u$, $\sigma_c$, $\sigma^{(r)}_c$
and $\sigma^{(r)}_u$, respectively.
We give an easier way to derive (\ref{chi0})
in the Appendix~\ref{appendix1}, where
integration over the $\alpha$ variables as those of Gaussian,
instead of minimizing with respect to
these variables, do the same job.
Eq.~(\ref{chi0}) shows that the square of the total systematic error is
given by the sum of the squares of all the systematic errors.

 Our strategy in this paper is to assume no neutrino oscillation
for the theoretical predictions $t$'s and assume
the number of events with oscillations for the measured values $m$'s.
 Then,
we examine whether a hypothesis with no
oscillation is excluded or not, say at the 90\%CL, from the value
of $\chi^2$.
 In the context of neutrino oscillation experiments,
we have
\begin{eqnarray}
\displaystyle
{m \over t}-1 = -\sin^22\theta
\left\langle \sin^2\left({\Delta m^2L \over 4E}
\right) \right\rangle
\label{cond0}
\end{eqnarray}
in the two flavor framework, where
$\theta$ is the mixing angle, $\Delta m^2$
is the mass squared difference%
\footnote{
Throughout this paper, we use the two flavor framework.
To translate it into the three flavor notation,
$\theta$ and $\Delta m^2$
should be interpreted as $\theta_{13}$
and $|\Delta m_{31}^2|$, respectively.},
$E$ is the neutrino energy,
$L$ is the distance of the reactor and the detector, and
\begin{eqnarray}
%\hspace*{-20mm}
\displaystyle
\left\langle \sin^2\left({\Delta m^2L_j \over 4E}
\right) \right\rangle\equiv
{\displaystyle
\int dE~\epsilon(E)f(E)\sigma(E)\sin^2\left({\Delta m^2L_j \over 4E}
\right) \over \displaystyle
\int dE~\epsilon(E)f(E)\sigma(E)}.
\nonumber
\end{eqnarray}
$\epsilon(E)$, $f(E)$, $\sigma(E)$ stand for the detection
efficiency, the neutrino flux, and the cross section, respectively.

\subsection{Two detectors}

Next let us discuss a less trivial example with a single
reactor, one near and one far detectors.
Let $m_{\text{n}}$ and $m_{\text{f}}$ be the measured numbers of events
at the near and far detectors, $t_{\text{n}}$ and $t_{\text{f}}$
be the theoretical predictions, respectively.
Then $\chi^2$ is given by
\begin{eqnarray}
\displaystyle
\chi^2&=&\min_{\alpha's}\left\{
\left[{m_{\text{n}}-t_{\text{n}}(1+\alpha_{\text{c}}
+\alpha_{\text{c}}^{\text{(r)}}
+\alpha_{\text{u}}^{(r)})
\over t_{\text{n}}\sigma_{\text{u}}}\right]^2
+\left[{m_{\text{f}}-t_{\text{f}}(1+\alpha_{\text{c}}
+\alpha_{\text{c}}^{\text{(r)}}
+\alpha_{\text{u}}^{(r)})
\over t_{\text{f}}\sigma_{\text{u}}}\right]^2
\right.\nonumber\\
&{\ }&\left.
+\left({\alpha_{\text{c}} \over \sigma_{\text{c}}}\right)^2
+\left({\alpha_{\text{c}}^{\text{(r)}} \over 
\sigma_{\text{c}}^{\text{(r)}}}\right)^2
+\left({\alpha_{\text{c}}^{\text{(r)}} \over 
\sigma_{\text{c}}^{\text{(r)}}}\right)^2\right\},
\label{chi1}
\end{eqnarray}
where we have assumed that the uncorrelated errors
for the two detectors are the same and are equal to
$\sigma_{\text{u}}$.
Eq.~(\ref{chi1}) can be evaluated also by
integrating over the variables $\alpha_{\text{c}}$, etc.\
as Gaussian
instead of minimizing with respect to these
variables.  After some calculations (See Appendix~\ref{appendix1}
for details), we obtain
\begin{eqnarray}
\hspace*{-20mm}
\displaystyle
\chi^2=
\left(\begin{array}{cc}
\displaystyle{m_{\text{n}} \over t_{\text{n}}}-1, &
\displaystyle{m_{\text{f}} \over t_{\text{f}}}-1
\end{array}
\right)
V^{-1}
\left(\begin{array}{c}
\displaystyle{m_{\text{n}} \over t_{\text{n}}}-1 \\
\\
\displaystyle{m_{\text{f}} \over t_{\text{f}}}-1
\end{array}
\right),
\nonumber
\end{eqnarray}
where
\begin{eqnarray}
V
&\equiv&
\sigma^2_{\text{u}}\, I_2
+ \left[
   \sigma^2_{\text{c}} + (\sigma^{\text{(r)}}_{\text{u}})^2
   + (\sigma^{\text{(r)}}_{\text{c}})^2
  \right] H_2\nonumber\\[3mm]
&=&
\left(
\displaystyle
\begin{array}{rr}
\sigma^2_{\text{u}}+\sigma^2_{\text{c}}
+(\sigma^{\text{(r)}}_{\text{u}})^2
+(\sigma^{\text{(r)}}_{\text{c}})^2\hspace*{5mm}
& \quad\sigma^2_{\text{c}}
+(\sigma^{\text{(r)}}_{\text{u}})^2
+(\sigma^{\text{(r)}}_{\text{c}})^2\\[3mm]
\quad\sigma^2_{\text{c}}
+(\sigma^{\text{(r)}}_{\text{u}})^2
+(\sigma^{\text{(r)}}_{\text{c}})^2\hspace*{5mm} &
\sigma^2_{\text{u}}+\sigma^2_{\text{c}}
+(\sigma^{\text{(r)}}_{\text{u}})^2
+(\sigma^{\text{(r)}}_{\text{c}})^2
\end{array}
\right)
\label{1r2d}
\end{eqnarray}
is the covariance matrix;
 $I_2$ represents $2\times 2$ identity matrix
and $H_2$ does a $2\times 2$ matrix whose elements are all unity.
 It is seen that
only the covariant matrix in the $\chi^2$ depends on the errors.
 Note that any liner transformation of $V$
does not change the value of $\chi^2$.
 Diagonalization of $V$ is, however, worthwhile
to investigate analytically the behavior of $\chi^2$.
After diagonalizing $V$ we have
\begin{eqnarray}
\hspace*{-20mm}
\displaystyle
\chi^2
&=&
{\left[\left(m_{\text{n}}/t_{\text{n}}-1\right)
+\left(m_{\text{f}}/t_{\text{f}}-1\right)
\right]^2 
\over 4\sigma^2_{\text{c}}
+4(\sigma^{\text{(r)}}_{\text{u}})^2
+4(\sigma^{\text{(r)}}_{\text{c}})^2
+2\sigma^2_{\text{u}}}
+{\left[\left(m_{\text{n}}/t_{\text{n}}-1\right)
-\left(m_{\text{f}}/t_{\text{f}}-1\right)
\right]^2 
\over 2\sigma^2_{\text{u}}}
\label{chi2}\\[3mm]
&=&
\sin^42\theta\left[
{\left(D(L_{\text{f}})+D(L_{\text{n}})\right)^2 \over
4\sigma^2_{\text{c}}+4(\sigma^{\text{(r)}}_{\text{u}})^2
+4(\sigma^{\text{(r)}}_{\text{c}})^2
+2\sigma^2_{\text{u}}}
+{\left(D(L_{\text{f}})-D(L_{\text{n}})\right)^2 \over 
2\sigma^2_{\text{u}}}\right],
\label{chi3}
\end{eqnarray}
where $L_{\text{n}}$ and $L_{\text{f}}$ are the distances from the reactor
to the near and far detector, respectively;
 We have defined
\begin{eqnarray}
\displaystyle
D(L)\equiv\left\langle \sin^2\left({\Delta m^2L \over 4E}
\right) \right\rangle.
\label{d}
\end{eqnarray}
 The first term on the right hand side in Eq.~(\ref{chi2})
stands for the contribution from the sum of the yields
at the near and far detectors, while the second term
corresponds to the difference between them.
 The first term determines the normalization of flux,
namely the sensitivity on $\sin^2{2\theta}$
at very large $|\Delta m^2|$
where all $D(L)$ becomes $0.5 \sin^2{2\theta}$.
 On the other hand,
the second term gives the main sensitivity
at the concerned value of $|\Delta m^2|$
(e.g.\ $2.5\times 10^{-3}\eV^2$)
as we see below.

Putting the reference values (\ref{error1}) and (\ref{error2})
together, we have
\begin{eqnarray}
\hspace*{-20mm}
2\sigma_{\text{u}}^2&=&(0.8\%)^2,\nonumber\\
4\sigma_{\text{c}}^2+4(\sigma^{\text{(r)}}_{\text{u}})^2
+4(\sigma^{\text{(r)}}_{\text{c}})^2+2\sigma_{\text{u}}^2&=&
(7.6\%)^2.\nonumber
\nonumber
\end{eqnarray}
 We can ignore the contribution from
$(4\sigma_{\text{c}}^2
+4(\sigma^{\text{(r)}}_{\text{u}})^2
+4(\sigma^{\text{(r)}}_{\text{c}})^2
+2\sigma_{\text{u}}^2)^{-1}$
in Eq.~(\ref{chi3})
because that is only 1\% compared to that from
$(2\sigma_{\text{u}}^2)^{-1}$.%
\footnote{
This is more or less the derivation
of $\chi^2$ used in \cite{Minakata:2002jv}.}
Hence, $\chi^2$ is given approximately by
\begin{eqnarray}
\chi^2\simeq\sin^42\theta
{\left(D(L_{\text{f}})-D(L_{\text{n}})\right)^2 \over 
2\sigma^2_{\text{u}}}.
\label{chi3.5}
\end{eqnarray}
 We see that the main sensitivity is determined indeed
by the relative normalization error
$\sigma_{\text{rel}} = \sqrt{2} \sigma_{\text{u}}$.
 By comparing (\ref{chi0}) and (\ref{chi3.5}),
it is clear that the sensitivity is improved significantly
by virtue of near detector.
 The hypothesis of no oscillation is excluded at the 90\%CL
if $\chi^2$ is larger than 2.7,
which corresponds to the value at the 90\%CL for one degree of freedom.
 This implies that the systematic limit
on $\sin^22\theta$ at the 90\%CL, namely the sensitivity
in the limit of infinite statistics, is given by
\begin{eqnarray}
\left(\sin^22\theta\right)_{\text{limit}}^{\text{sys~only}}
\simeq\sqrt{2.7}{\sqrt{2}\sigma_{\text{u}} \over 
D(L_{\text{f}})-D(L_{\text{n}})}.
\label{sens0}
\end{eqnarray}
 Eq.~(\ref{sens0}) also tells us that, in order to
optimize $\left(\sin^22\theta\right)_{\text{limit}}^{\text{sys~only}}$
for a given value of $\sigma_{\text{u}}$,
we have to maximize $D(L_{\text{f}})\equiv
\langle \sin^2\left({\Delta m^2L_{\text{f}} / 4E}
\right)\rangle$ while minimizing $D(L_{\text{n}})\equiv
\langle \sin^2\left({\Delta m^2L_{\text{n}} / 4E}
\right)\rangle$.
 Note that $D(L_{\text{f}})$ can not be unity
because of neutrino energy spectrum;
 The possible maximum value of
$D(L_{\text{f}})-D(L_{\text{n}})$ is 0.82,
which is attained for $\Delta m^2 = 2.5\times 10^{-3}\eV^2$,
$L_{\text{f}}=1.8$km, and $L_{\text{n}}=0$.
 Then,
we can estimate the
lower bound of
$\left(\sin^22\theta\right)_{\text{limit}}^{\text{sys~only}}$
at a single reactor experiment,
assuming that the uncorrelated error $\sigma_{\text{u}}$
is smaller enough than other errors:
\begin{eqnarray}
\left(\sin^22\theta\right)_{\text{limit}}^{\text{sys~only}}
\gtrsim{\sqrt{2.7}\sqrt{2} \sigma_{\text{u}}\over 0.82}
=2.8\,\sigma_{\text{u}}.
\label{sens1}
\end{eqnarray}
 In practice, however,
$D(L_{\text{n}})$ will not be able to vanish.
 Assuming $\Delta m^2=2.5\times10^{-3}\eV^2$,
$L_{\text{f}}=1.7\km$, and $L_{\text{n}}=0.3\km$,
we have $D(L_{\text{f}}) \simeq 0.82$
and $D(L_{\text{n}}) \simeq 0.07$.
 Then,
$\sigma_{\text{u}}=(0.8/\sqrt{2})\%$ gives the sensitivity
\begin{eqnarray}
\left(\sin^22\theta\right)_{\text{limit}}^{\text{sys~only}}\simeq
\sqrt{2.7}{\sqrt{2}\sigma_{\text{u}} \over D(L_{\text{f}})-D(L_{\text{n}})}
\simeq 3.1\sigma_{\text{u}} \simeq 0.018.
\nonumber
\end{eqnarray}
 This sensitivity corresponds to and agrees with
the value obtained numerically in \cite{Minakata:2002jv}.

\section{$n_r$ reactors}
It is straightforward to generalize the argument in the
previous section to a general case with multi-reactors
and multi-detectors.  The covariance matrix
$V$ is given by $\sigma_{\text{u}}^2\times$(unit matrix)%
+(the rest), and in most cases, as long as the near detectors are
placed properly, the determinant of (the
rest) is zero or very small compared to $\sigma_{\text{u}}^2$.
The minimum eigenvalue of the covariance matrix, which gives main
contribution to $\chi^2$, is approximately given by $\sigma_{\text{u}}^2$.
 Therefore,
the systematic limit is dominated by the uncorrelated error
$\sigma_{\text{u}}$ also in general cases.

\subsection{One detector}
As in Sect.~\ref{onereactor}, as a warming up,
let us consider the case with one
detector and multiple reactors (See Fig.\ref{fig1}(a)).
When there are $n_r(>1)$ reactors, the total number $m$ of the
measured events
is a sum of contributions $m_a~(a=1,\cdots,n_r)$ from
each reactor, and this is also the case for the
theoretical predictions $t$ and $t_a~(a=1,\cdots,n_r)$.
So we have
\begin{eqnarray}
\hspace*{-20mm}
m=\sum_{a=1}^{n_r} m_a,\quad
t=\sum_{a=1}^{n_r} t_a.
\nonumber
\end{eqnarray}
Taking these systematic errors into consideration,
the total number $t$ of the theoretical prediction
is redefined as
\begin{eqnarray}
\hspace*{-20mm}
t&\rightarrow& \sum_{a=1}^{n_r} t_a\left(1+\alpha_{\text{c}}
+\alpha_{\text{c}}^{\text{(r)}}
+\alpha_{\text{u}a}^{(r)}\right)\nonumber\\
&=&t\left(1+\alpha_{\text{c}}+\alpha_{\text{c}}^{\text{(r)}}\right)
+\sum_{a=1}^{n_r} t_a \alpha_{\text{u}a}^{(r)}
=t\left(1+\alpha_{\text{c}}
+\alpha_{\text{c}}^{\text{(r)}}
+\sum_{a=1}^{n_r} {t_a \over t} \alpha_{\text{u}a}^{(r)}\right),
\nonumber
\end{eqnarray}
where $\alpha_{\text{u}a}^{(r)}$ is the variable to
introduce the uncorrelated of the flux from the $a$-th
reactor.  $\chi^2$ is defined as
\begin{eqnarray}
\hspace*{-20mm}
\chi^2&=&\min_{\alpha's}\left\{
{1 \over t^2\sigma^2}
\left[m-t\left(1+\alpha_{\text{c}}+\alpha_{\text{c}}^{\text{(r)}}
+\sum_{a=1}^{n_r} {t_a \over t} \alpha_{\text{u}a}^{\text{r}}
\right)\right]^2
\right.\nonumber\\
&{\ }&\left.
+\left({\alpha_{\text{c}} \over \sigma_{\text{c}}}\right)^2
+\left({\alpha_{\text{c}}^{\text{(r)}} \over 
\sigma_{\text{c}}^{\text{(r)}}}\right)^2
+\sum_{a=1}^{n_r} \left({\alpha_{\text{u}a}^{\text{(r)}} \over 
\sigma_{\text{u}}^{\text{(r)}}}\right)^2\right\},
\nonumber
\end{eqnarray}
where we have again assumed for simplicity that the size of the uncorrelated
error in the flux from the reactors is common:
$\sigma_{\text{u}a}^{\text{(r)}} = \sigma_{\text{u}}^{\text{(r)}}$.
Performing minimization with respect to the $\alpha$ variables,
we get (See Appendix~\ref{appendix1})
\begin{eqnarray}
\hspace*{-20mm}
\chi^2
={\left(\displaystyle{m \over t}-1\right)^2 \over
\displaystyle \sigma_{\text{u}}^2+\sigma_{\text{c}}^2
+(\sigma_{\text{c}}^{\text{(r)}})^2
+(\sigma_{\text{u}}^{\text{(r)}})^2
\sum_{a=1}^{n_r} \left({t_a \over t}\right)^2}.
\label{chi4}
\end{eqnarray}
 By comparing (\ref{chi4}) with (\ref{chi0}),
we find that the multiple reactor nature
affects $\chi^2$ (the sensitivity on $\sin^22\theta$)
through $\sigma_{\text{u}}^{\text{(r)}}$.
Since there is only one detector in this case,
the correlated error was not canceled in $\chi^2$
and it contributes to the systematic limit on $\sin^22\theta$.
However, if the yield from each reactor is approximately equal,
i.e., if
\begin{eqnarray}
{t_a \over t}\simeq {1 \over n_r},
\label{equal}
\end{eqnarray}
then the contribution of the uncorrelated
error of the reactors becomes
\begin{eqnarray}
(\sigma_{\text{u}}^{\text{(r)}})^2
\sum_{a=1}^{n_r} \left({t_a \over t}\right)^2
\simeq\displaystyle{1 \over n_r}(\sigma_{\text{u}}^{\text{(r)}})^2.
\label{reduction}
\end{eqnarray}
Comparing Eqs.~(\ref{chi0}) and (\ref{chi4}), we observe
that the contribution of the uncorrelated
error of the reactors decreases as the number of the reactors
increases, as long as the condition (\ref{reduction}) is satisfied.%
\footnote{One can show from the Cauchy-Schwarz inequality
that $\sum_{a=1}^{n_r} \left({t_a/t}\right)^2\le 1$
always holds
even if the condition (\ref{reduction}) is not satisfied.
 Hence,
the contribution of the uncorrelated
error of the reactors decreases always.}
This is because the average of independent $n_r$ fluctuations
is smaller than a single fluctuation.

\subsection{$n_d$ detectors}
Let us now discuss more general cases with $n_r$ reactors
and $n_d$ detectors.  For simplicity we assume again that
the size of the uncorrelated errors for the detectors are the same and
the size of the uncorrelated errors in the flux from the reactors
are also the same:
$\sigma_{\text{u}j} = \sigma_{\text{u}}$,
$\sigma_{\text{u}a}^{(\text{r})} = \sigma_{\text{r}}^{\text{u}}$.
Let $t_{ja}$ ($m_{ja}$) be the theoretical prediction 
(measured value) for
the number of events of neutrinos from the $a$-th reactor $(a=1,\cdots,n_r)$
at the $j$-th detector $(j=1,\cdots,n_d)$ and
$t_j=\sum_{a=1}^{n_r} t_{ja}$ ($m_j=\sum_{a=1}^{n_r} m_{ja}$)
be the theoretical (measured) total number of events at the $j$-th detector.
Then generalizing the discussions
in the previous sections, we have
\begin{eqnarray}
\hspace*{-20mm}
\chi^2&=&\min_{\alpha's}\left\{
\displaystyle\sum_{j=1}^{n_d}
{1 \over t_j^2\sigma_{\text{u}}^2}
\left[m_j-t_j\left(1+\alpha_{\text{c}}+\alpha_{\text{c}}^{\text{(r)}}
+\sum_{a=1}^{n_r}
{t_{aj} \over t_j} \alpha_{\text{u}a}^{(r)}\right)\right]^2\right.
\nonumber\\
&{\ }&+\left.\left({\alpha_{\text{c}} \over \sigma_{\text{c}}}\right)^2
+\left({\alpha_{\text{c}}^{\text{(r)}} \over 
\sigma_{\text{c}}^{\text{(r)}}}\right)^2
+\sum_{a=1}^{n_r} \left({\alpha_{\text{u}a}^{\text{(r)}} \over 
\sigma_{\text{u}}^{\text{(r)}}}\right)^2\right\}.
\nonumber
\end{eqnarray}
After some calculation (See Appendix~\ref{appendix1}), we have
\begin{eqnarray}
\chi^2
&=&\left(\displaystyle{m_1 \over t_1}-1,\cdots,
\displaystyle{m_{n_d} \over t_{n_d}}-1\right)
V^{-1}
\left(
\begin{array}{c}
\displaystyle{m_1 \over t_1}-1\\
\vdots\\
\displaystyle{m_{n_d} \over t_{n_d}}-1
\end{array}
\right),
\nonumber
\end{eqnarray}
where
\begin{eqnarray}
V_{jk}&=&\delta_{jk}\sigma_{\text{u}}^2
+\sigma_{\text{c}}^2
+(\sigma_{\text{c}}^{\text{(r)}})^2
+(\sigma_{\text{u}}^{\text{(r)}})^2\displaystyle 
\sum_{a=1}^{n_r} \displaystyle{t_{aj} \over t_j}
{t_{ak} \over t_k}.
\label{chi5}
\end{eqnarray}

\subsubsection{$n_r$ reactors and $(n_r+1)$ detectors\label{sssect:1r1n}}
In Sect.~\ref{onereactor} we have seen that the correlated error
is canceled in the case of a single reactor
experiment with one near and one far detectors.
Now we would like to ask the following question:
what happens to this cancellation in the case of
an experiment with multiple reactors and detectors?
To answer this question, let us consider the ideal
case with $n_r$ reactors and $(n_r+1)$ detectors,
where each reactor has a near detector in its neighborhood
and each reactor produces the same number of events at a far detector
(See Fig.\ref{fig1}(b)):
\begin{eqnarray}
\mbox{\rm near detectors}:~&{\ }&
\displaystyle{t_{aj} \over t_j}=\delta_{aj}~\qquad\quad
(j=1,\cdots,n_r;a=1,\cdots,n_r)
\label{frac1}\\
\mbox{\rm far detector}:~&{\ }&
\displaystyle{t_{a\,n_r+1} \over t_{n_r+1}}=\displaystyle{1 \over n_r}\qquad
(a=1,\cdots,n_r).
\nonumber
\end{eqnarray}
In this case the element of the covariance matrix becomes
\begin{eqnarray}
\hspace*{-10mm}
V&=&\sigma_{\text{u}}^2\, I_{n_r+1}
+\left[\sigma_{\text{c}}^2
+\left(\sigma_{\text{c}}^{(r)}\right)^2\right]H_{n_r+1}
+\left(\sigma_{\text{u}}^{(r)}\right)^2
\left(\begin{array}{ccccc}
1     &0     &\cdots&0     &1/n_r\\
0     &\ddots&      &\vdots&\vdots\\
\vdots&      &\ddots&0     &\vdots\\
0     &\cdots&0     &1     &\vdots\\[3mm]
1/n_r &\cdots&\cdots&\cdots&1/n_r
\end{array}\right),
\label{vnrnr1d}
\end{eqnarray}
where $I_{n_r+1}$ is an $(n_r+1)\times (n_r+1)$ unit matrix,
and $H_{n_r+1}$ is an $(n_r+1)\times (n_r+1)$ matrix defined by
\begin{eqnarray}
\hspace*{-20mm}
H_{n_r+1}&\equiv&\left(\begin{array}{ccc}
1&\cdots&1\\
\vdots&&\vdots\\
1&\cdots&1
\end{array}\right).
\label{h}
\end{eqnarray}
Here we assume the following conditions\footnote{
For simplicity we assume here that the distance
between the $a$-th reactor and its near detector
is equal to $L_{\text{n}}$ for $a=1,\cdots,n_r$.
In order for (\ref{frac1}) to be satisfied,
Eq.~(\ref{cond1}) is necessary.  So in this ideal situation
which we are considering, the dependence on
$\left\langle\sin^2\left(
{\Delta m^2L_{\text{n}} / 4E}\right)
\right\rangle$ cannot be discussed in a manner
consistent with the assumption (\ref{frac1}).}:
\begin{eqnarray}
\hspace*{-20mm}
\left| {\Delta m^2 L_{\text{n}} \over 4E}\right| \ll 1, \ \ \
\left| {\Delta m^2 L_{\text{f}} \over 4E}\right| \simeq {\pi \over 2}.
\label{cond1}
\end{eqnarray}
 These conditions have to be satisfied
in an experiment which aims to measure $\theta_{13}$.
From Eqs.~(\ref{cond1}) we have
\begin{eqnarray}
\left(\begin{array}{c}
m_1/t_1-1\\
\vdots\\
m_{n_r}/t_{n_r}-1\\
m_{n_r+1}/t_{n_r+1}-1\\
\end{array}
\right)
\simeq\left(\begin{array}{c}
0\\
\vdots\\
0\\
-\sin^22\theta\left\langle\sin^2\left(
\displaystyle{\Delta m^2L_{\text{f}} \over 4E}\right)
\right\rangle\\
\end{array}\right)\equiv-\sin^22\theta D(L_{\text{f}})
\left(\begin{array}{c}
0\\
\vdots\\
0\\
1
\end{array}\right)
,
\nonumber
\end{eqnarray}
and $\chi^2$ becomes (See appendix \ref{appendix2} for derivation.)
\begin{eqnarray}
\chi^2=\sin^42\theta D(L_{\text{f}})^2
\left\{{1 \over n_r+1}{1 \over \sigma_{\text{u}}^2
+(n_r+1)\left[\sigma_{\text{c}}^2
+\left(\sigma_{\text{c}}^{(r)}\right)^2
+\left(\sigma_{\text{u}}^{(r)}\right)^2/n_r
\right]}+{n_r \over n_r+1}{1 \over \sigma_{\text{u}}^2}
\right\}.
\label{chinrnr1d}
\end{eqnarray}
 Then, the systematic limit on $\sin^22\theta$ at the 90\%CL
is given by
\begin{eqnarray}
&&\hspace*{-5mm}
\left(\sin^22\theta\right)_{\text{limit}}^{\text{sys~only}}\nonumber\\[-3mm]
&&
=\sqrt{2.7}\sqrt{1+\displaystyle{1 \over n_r}}
\,{\sigma_{\text{u}} \over D(L_{\text{f}})}
\,\left\{1+{\sigma_{\text{u}}^2/n_r \over \sigma_{\text{u}}^2
+(n_r+1)\left[\sigma_{\text{c}}^2
+\left(\sigma_{\text{c}}^{(r)}\right)^2
+\left(\sigma_{\text{u}}^{(r)}\right)^2/n_r
\right]}\right\}^{-\displaystyle{1 \over 2}}\label{sigmaeff2.5}\\
&&
\simeq\sqrt{2.7}\sqrt{1+\displaystyle{1 \over n_r}}
\,{\sigma_{\text{u}} \over D(L_{\text{f}})}.
\label{sigmaeff3}
\end{eqnarray}
 For example,
we obtain
$\left(\sin^22\theta\right)_{\text{limit}}^{\text{sys~only}} \simeq
0.012$
by assuming seven reactors, $L_{\text{f}} = 1.7\km$,
and $\sigma_{\text{u}} = 0.8/\sqrt{2} \%$.
As in the case with one reactor,
the dominant contribution to the systematic limit comes
from the uncorrelated error $\sigma_{\text{u}}$.
 The contribution of the uncorrelated 
error of flux, $\sigma^{(r)}_{\text{u}}$, is reduced in (\ref{sigmaeff2.5})
by a factor of
$n_r$ due to the averaging over the independent $n_r$
fluctuations;
 Although this reduction is a potential merit
of the multiple reactor complex,
it is irrelevant to the sensitivity because such
an effect comes in the correlated error among detectors which is
almost canceled in the multi-detector system.
The factor $\sqrt{1+1/n_r}$ which appears in the
dominant contribution by $\sigma_{\text{u}}$
indicates that the effective systematic error
decreases as the number $(1+n_r)$ of the detectors
increases, since more information is obtained with
more detectors.
To conclude, the answer to the question at the
beginning of this subsection is that
the cancellation of the correlated error occurs
also in the ideal case with $n_r$ reactors and $(n_r+1)$ detectors,
and once again the systematic limit is dominated by
$\sigma_{\text{u}}$.
It should be noted that the number $n_r$ of the near
detectors in this case is sufficient but not necessary
to guarantee this cancellation, as we will see below in the
case of the KASKA plan.

\subsubsection{The case of the KASKA plan}
\begin{figure}
\includegraphics[scale=0.32]{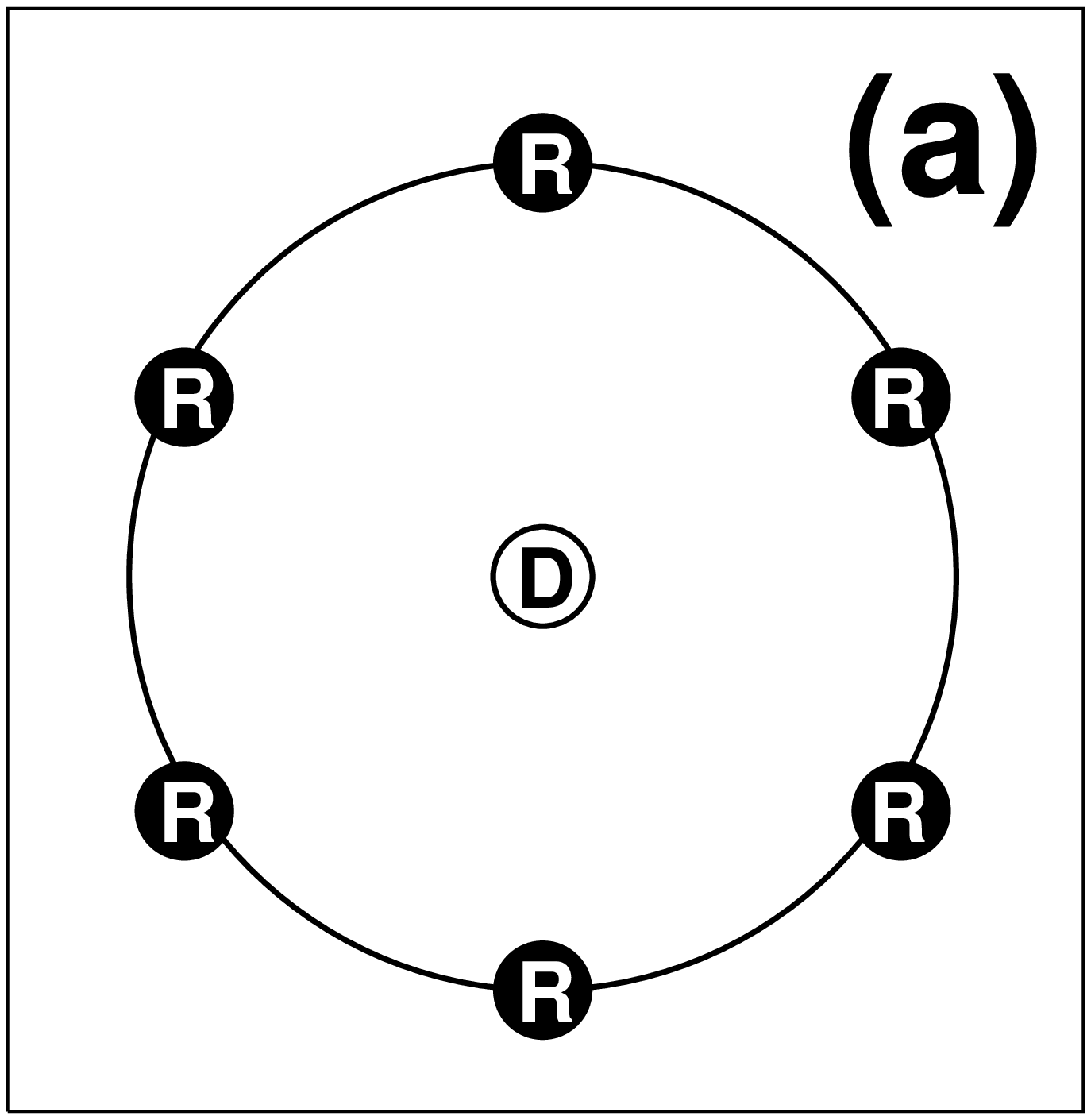}
\includegraphics[scale=0.32]{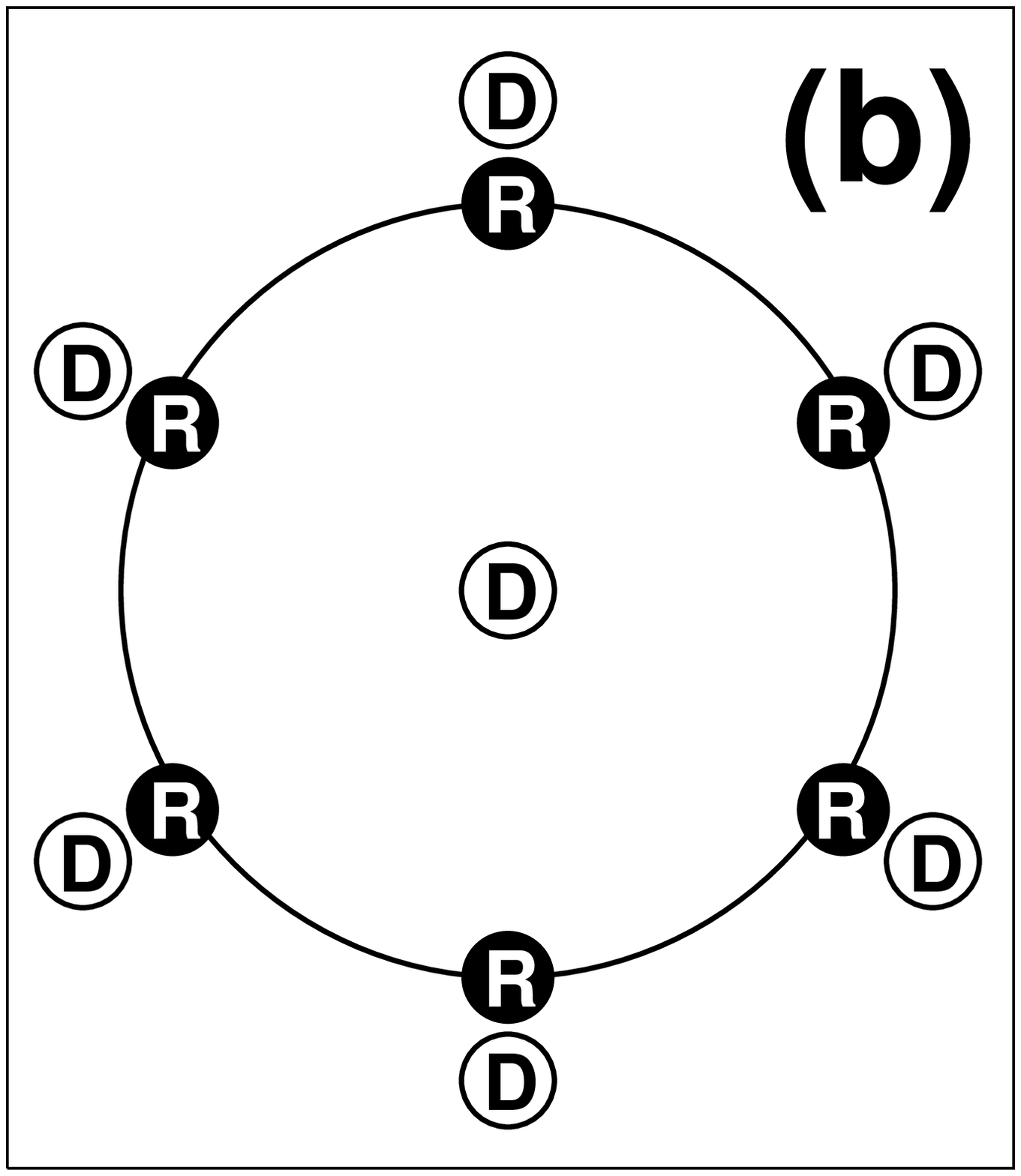}
\includegraphics[scale=0.32]{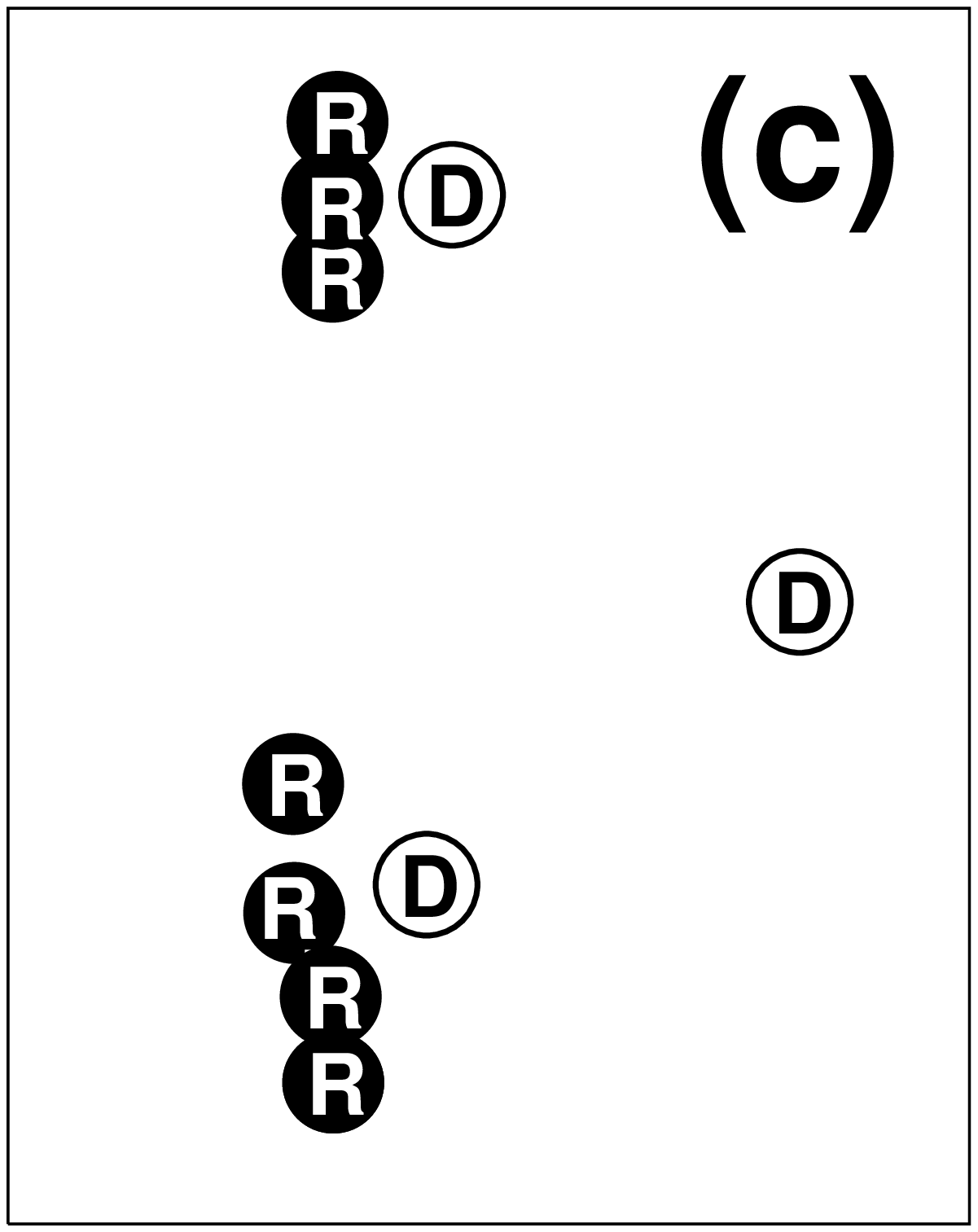}
\includegraphics[scale=0.32]{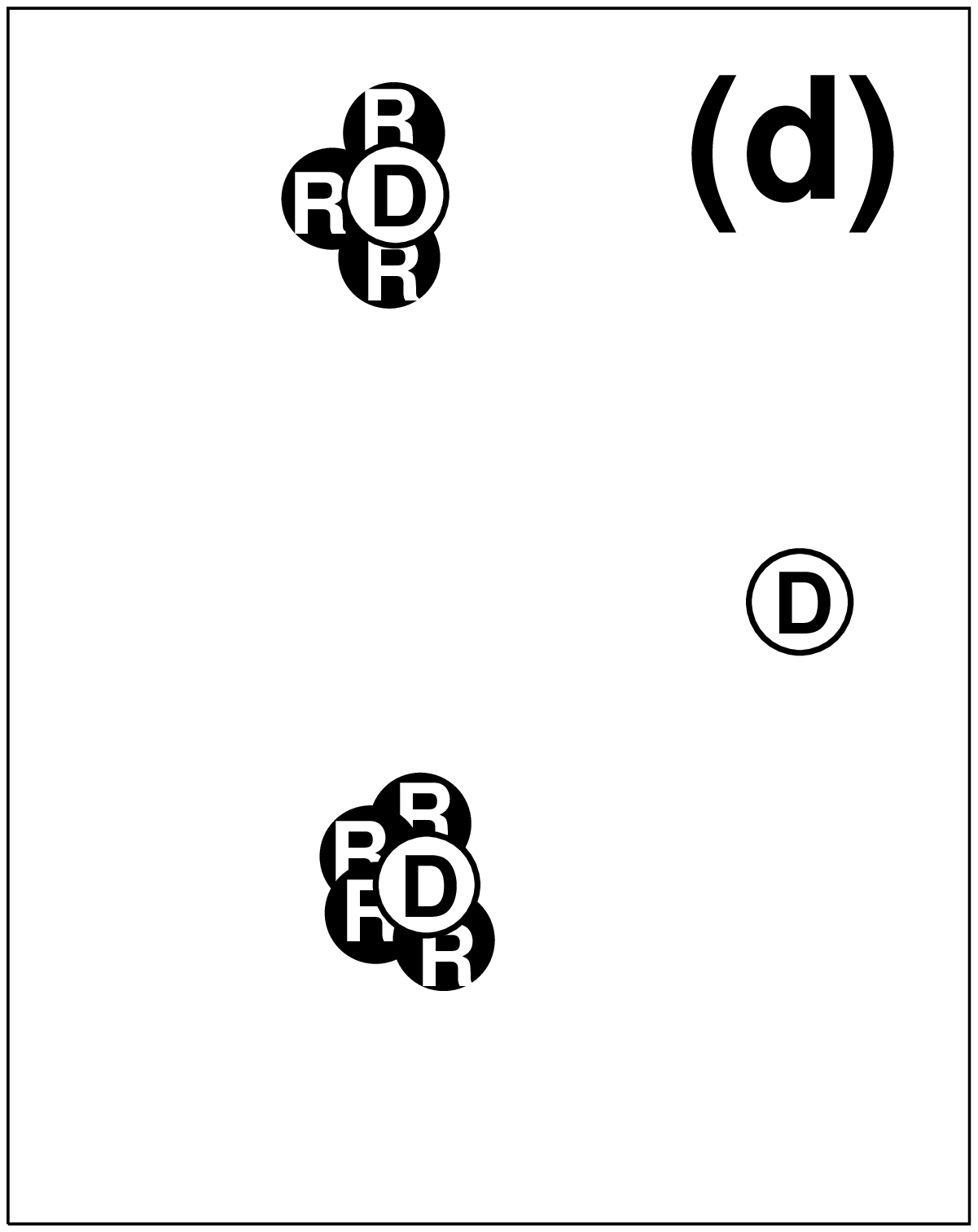}
\includegraphics[scale=0.32]{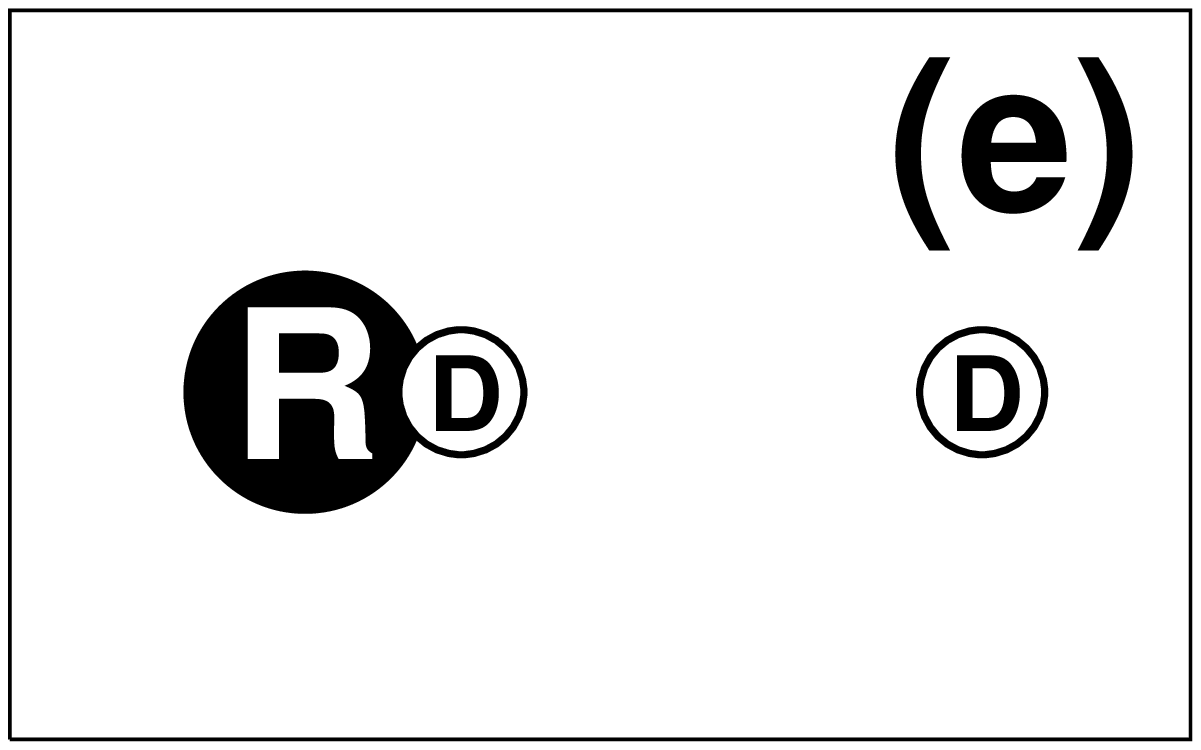}
\includegraphics[scale=0.32]{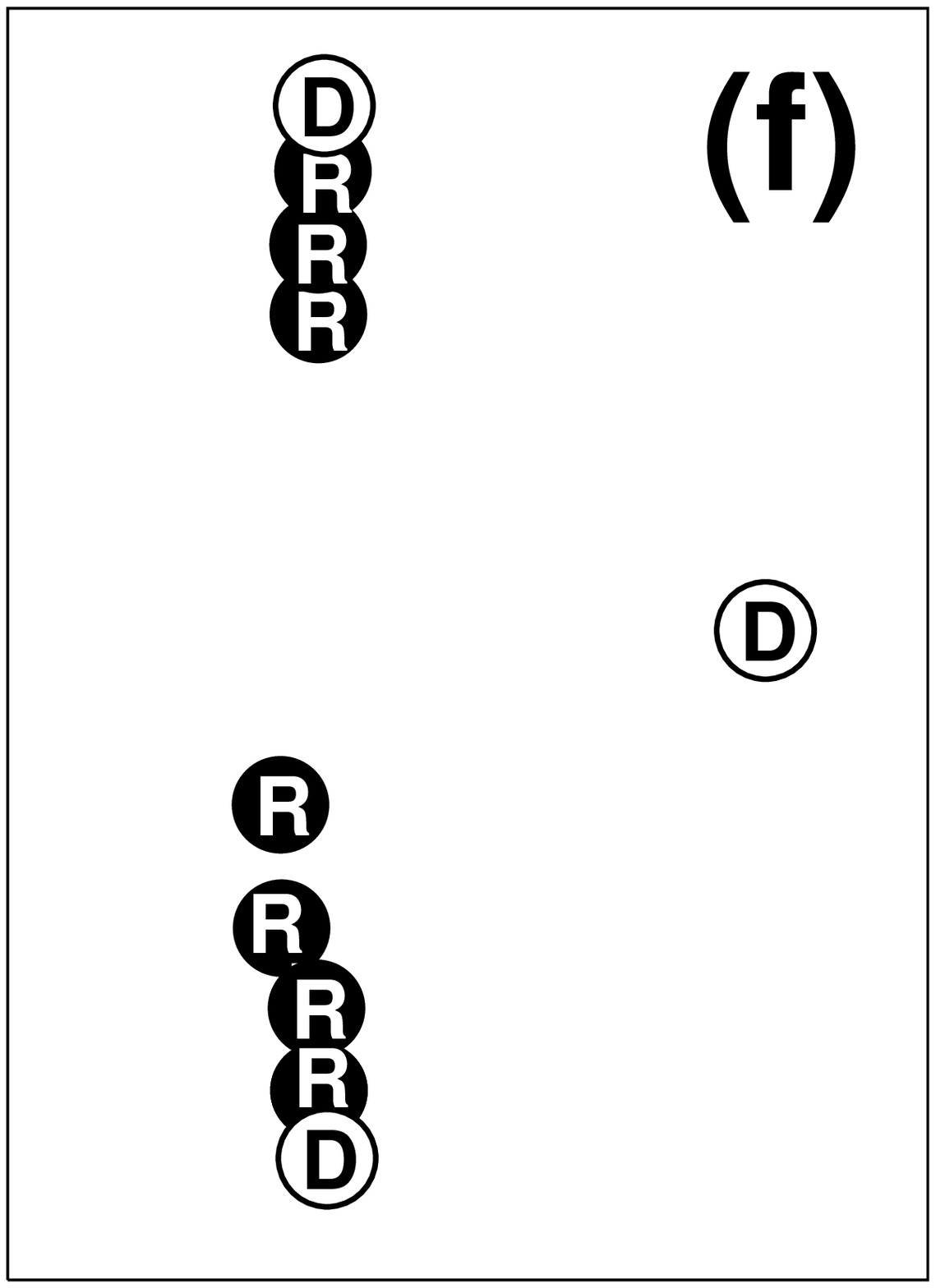}
\vglue 0.5cm
\caption{
Configurations of the experiments:
(a) $n_r$ reactors + one detector.
(b) $n_r$ reactors + $(n_r+1)$ detectors.
(c) The KASKA plan with 7 reactors + 3 detectors.
(d) The ideal limit of the KASKA plan.
(e) One reactor + 2 detectors to be compared with (c).
(f) The hypothetical KASKA plan with the near detectors
in the wrong locations to be compared with (c).
}
\label{fig1}
\end{figure}

The Kashiwazaki-Kariwa nuclear power station consists of two clusters
of reactors, and one cluster consists of four reactors while the
other consists of three (See Fig.\ref{fig1}(c)).
 According to the discussion in the previous section,
we understand that near-far cancellation
occurs for the KASKA case
if we have seven near detectors.
 In the KASKA plan, however,
not each reactors but each cluster of reactors
assumed to have a near detector.
In this subsection we would like to clarify
the following questions
on the KASKA plan~\cite{kaska}:
(a) Is the number of near detectors sufficient
for the cancellation of the correlated error?
(b) What are the effects of multiple sources?
(c) Is the KASKA plan optimized with respect to
the sensitivity to $\sin^22\theta$?
Here we again assume that the size of the uncorrelated
error in the flux from the reactors is common and
the size of the uncorrelated errors of the three detectors are the same.

Before we discuss the
systematic limit for the actual KASKA plan,
let us consider the ideal limit, in which all the
reactors in each cluster shrinks to one point
as is shown in Fig.\ref{fig1}(d);
 The ideal limit is similar to the case discussed
in the section~\ref{sssect:1r1n},
namely the case of two reactors and three detectors.
In this ideal limit, we have
\begin{eqnarray}
\hspace*{-40mm}
\mbox{\rm near 1:}\qquad
\displaystyle{t_{a1} \over t_1}&=&
\left\{\begin{array}{cc}
\displaystyle\frac{1}{\,4\,}&\qquad(a=1,\cdots,4)\\[5mm]
0&\qquad(a=5,6,7)
\end{array}\right.
\label{cond21}\\[3mm]
\hspace*{-40mm}
\mbox{\rm near 2:}\qquad
\displaystyle{t_{a2} \over t_2}&=&
\left\{\begin{array}{cc}
0&\qquad(a=1,\cdots,4)\\[3mm]
\displaystyle\frac{1}{\,3\,}&\qquad(a=5,6,7)
\end{array}\right.
\label{cond22}\\[3mm]
\hspace*{-40mm}
\mbox{\rm far:}\qquad
\displaystyle{t_{a3} \over t_3}&=&
\displaystyle\frac{1}{\,7\,}\qquad(a=1,\cdots,7)
\label{cond23}
\end{eqnarray}
and the covariance matrix is given by
\begin{eqnarray}
\hspace*{-20mm}
V&=&
\sigma_{\text{u}}^2 I_3
+\left[\sigma_{\text{c}}^2
+\left(\sigma_{\text{c}}^{(r)}\right)^2\right] H_3
+\left(\sigma_{\text{u}}^{(r)}\right)^2
\left(
\begin{array}{ccc}
1/4&
0 &
1/7\\
0 &
1/3 &
1/7\\
1/7 &
1/7 &
1/7
\end{array}
\right)\nonumber\\
&=&\left(
\begin{array}{lll}
\sigma^2_{\text{c}}+(\sigma^{\text{(r)}}_{\text{c}})^2
+\displaystyle{(\sigma_{\text{u}}^{\text{(r)}})^2 \over 4}
+\sigma^2_{\text{u}}&
\quad\sigma^2_{\text{c}}+(\sigma^{\text{(r)}}_{\text{c}})^2 &
\quad\sigma^2_{\text{c}}+(\sigma^{\text{(r)}}_{\text{c}})^2
+\displaystyle{(\sigma_{\text{u}}^{\text{(r)}})^2 \over 7}\\[3mm]
\sigma^2_{\text{c}} +(\sigma^{\text{(r)}}_{\text{c}})^2 &
\quad\sigma^2_{\text{c}}+(\sigma^{\text{(r)}}_{\text{c}})^2
+\displaystyle{(\sigma_{\text{u}}^{\text{(r)}})^2 \over 3}
+\sigma^2_{\text{u}} &
\quad\sigma^2_{\text{c}} +(\sigma^{\text{(r)}}_{\text{c}})^2
+\displaystyle{(\sigma_{\text{u}}^{\text{(r)}})^2 \over 7}\\[3mm]
\sigma^2_{\text{c}}+(\sigma^{\text{(r)}}_{\text{c}})^2
+\displaystyle{(\sigma_{\text{u}}^{\text{(r)}})^2 \over 7} &
\quad\sigma^2_{\text{c}}+(\sigma^{\text{(r)}}_{\text{c}})^2
+\displaystyle{(\sigma_{\text{u}}^{\text{(r)}})^2 \over 7} &
\quad\sigma^2_{\text{c}}+(\sigma^{\text{(r)}}_{\text{c}})^2
+\displaystyle{(\sigma_{\text{u}}^{\text{(r)}})^2 \over 7}
+\sigma^2_{\text{u}}
\end{array}
\right),\nonumber\\
\label{v7r3d}
\end{eqnarray}
where $I_3$ is a $3\times3$ unit matrix and $H_3$ is 
a $3\times3$ matrix defined in Eq.~(\ref{h}).
After diagonalizing $V$,
we obtain the systematic limit (See appendix\ref{appendix3}.)
\begin{eqnarray}
\left(\sin^22\theta\right)_{\text{limit}}^{\text{sys~only}}\simeq
\displaystyle{\sqrt{2.7} \over D(L_{\text{f}})}\,
\displaystyle{\sqrt{74} \over 7}\,
\sigma_{\text{u}}.
\label{limit1}
\end{eqnarray}
$\sigma_{\text{u}}$ is
the dominant contribution to the systematic limit
in Eq.~(\ref{limit1}) and the correlated errors are
canceled due to the near-far detector complex.
The reason that we have the factor $\sqrt{74/49}$ instead of
$(1+1/N)^{1/2}|_{N=2}=\sqrt{3/2}$ is because
the ratio of the $\overline{\nu}_e$ yield at the first cluster
to that at the second one is 4:3 instead of 1:1
assumed in (\ref{frac1}).

In reality, however, the conditions (\ref{cond21})--(\ref{cond23})
are not exactly satisfied in the setting
of the actual KASKA plan~\cite{kaska}.
Let us evaluate the exact eigenvalues of the
covariance matrix by taking into account the
actual parameters in \cite{kaska}.
Table \ref{table1} shows the power of the reactors
and the distance between the seven reactors and
the three detectors.  From this we can calculate
the fraction $t_{aj}/t_j~(a=1,\cdots,7,~j=1,2,3)$
which is given in Table~\ref{table2}.
The covariance matrix is now given by
\begin{eqnarray}
\hspace*{-10mm}
V&=&\sigma_{\text{u}}^2 I_3
+\left[\sigma_{\text{c}}^2
+\left(\sigma_{\text{c}}^{(r)}\right)^2\right] H_3
+\left(\sigma_{\text{u}}^{(r)}\right)^2
\left(
\begin{array}{lll}
 0.221 & 0.042 & 0.149\\
 0.042 & 0.298 & 0.138\\
 0.149 & 0.138 & 0.145
\end{array}
\right).
\label{v2}
\end{eqnarray}
Diagonalization of $V$ can be done only numerically
with the reference values in Eqs.~(\ref{error1}) and
(\ref{error2}).
We find that
the minimum eigenvalue of $V$ is $1.04\times\sigma_{\text{u}}^2$.
 The eigenvalue shows that the cancellation of the correlated errors
occurs also in the actual KASKA plan
though the number of near detectors is less than
that of reactors.
The systematic limit on $\sin^22\theta$ is approximately
given by the contribution from the minimum eigenvalue
(See appendix\ref{appendix3}.):
\begin{eqnarray}
\left(\sin^22\theta\right)_{\text{limit}}^{\text{sys~only}}
\simeq
{\sqrt{2.7}\times\sqrt{1.04}\sigma_{\text{u}} \over
\left|\overline{D(L_3)}u^{(1)}_3+
\overline{D(L_1)}u^{(1)}_1+
\overline{D(L_2)}u^{(1)}_2
\right|} \simeq 3.9\sigma_{\text{u}} \simeq0.022,
\label{limitkk}
\end{eqnarray}
where $\vec{u}^{(1)}\equiv(u^{(1)}_1,u^{(1)}_2,u^{(1)}_3)^T$
is the eigenvector of $V$ which corresponds to the
minimum eigenvalue ($\simeq \sigma_{\text{u}}^2$) of $V$,
and $\overline{D(L_j)}$ is the average of each
contribution $D(L_{aj})$:
\begin{eqnarray}
\overline{D(L_j)}\equiv
\displaystyle\sum_{a=1}^{7}\displaystyle{t_{aj} \over t_j}
\left\langle
\sin^2\left(
{\Delta m^2L_{aj} \over 4E}
\right)
\right\rangle\qquad (j=1(\mbox{\rm near}),~2(\mbox{\rm near}),
~3(\mbox{\rm far})).
\label{overlined}
\end{eqnarray}
Here $L_{aj}$ is the distance between the $a$-th reactor and
the $j$-th detector, and $t_{aj}/t_j$
is the fraction of the yield from the $a$-th reactor at
the detector $j$=1,~2,~3.
 When the near-far cancellation occurs sufficiently,
the value of
$\left(\sin^22\theta\right)_{\text{limit}}^{\text{sys~only}}/\sigma_{\text{u}}$
gives a good measure for the power of a reactor experiment
almost independently of assumptions of error sizes;
 The smaller value means the better setup of reactor experiments.

\begin{figure}
\hglue -1.8cm
\includegraphics[scale=0.6]{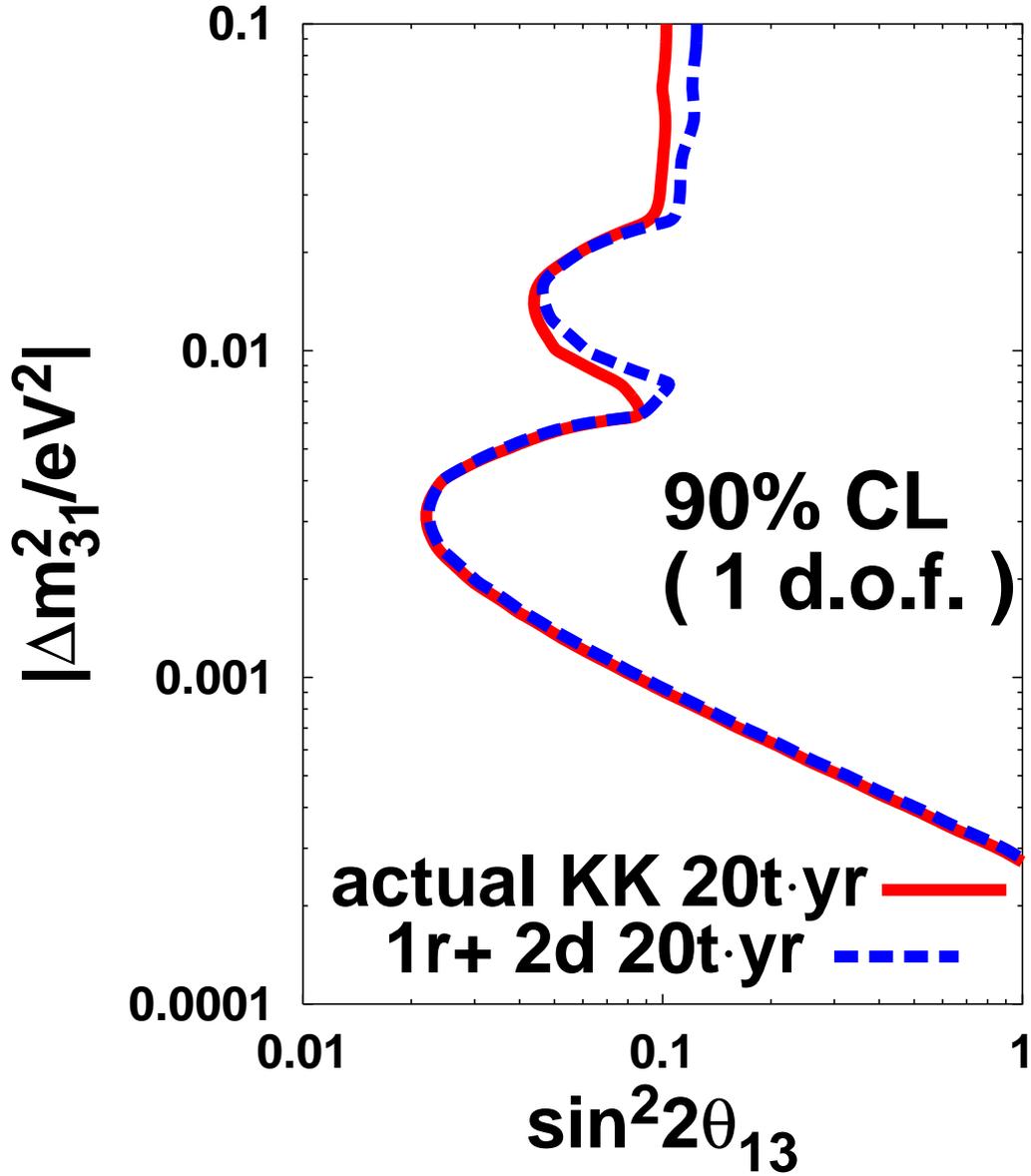}
\vglue 0.5cm
\caption{
The comparison of the sensitivity to $\sin^22\theta_{13}$
for the actual KASKA plan
(the solid line; with the configuration depicted in Fig.\ref{fig1}(c))
and a hypothetical case
(the dashed line; with the configuration depicted in Fig.\ref{fig1}(e)).
The statistical errors as well as all the systematic errors are taken into
account.}
\label{fig2}
\end{figure}

To see how effectively the correlated errors
are canceled in the actual KASKA plan,
comparison is given in Fig.\ref{fig2} between
the sensitivities to $\sin^22\theta$ of
the actual KASKA plan (Fig.\ref{fig1}(c))
and of a hypothetical experiment
with a single reactor and two detectors (300m and 1.3km baselines)
depicted in Fig.\ref{fig1}(e));
 The same value of uncorrelated systematic error
$\sigma_u$ and the same data size (=20t$\cdot$yr) are used
for each case.
 We observe that there is little difference between the sensitivities
at $\Delta m^2=2.5\times10^{-3}\eV^2$.
Also it is remarkable that the sensitivity of the actual KASKA plan
for higher value of $\Delta m^2$ is better than of the
single reactor experiment.  This is exactly because of the
reduction of the uncorrelated error from 
due to the nature of multi reactors
(cf.\ Eq.~(\ref{reduction})), where the near detectors
play a role as far detectors in this case.
 Here, it should be mentioned that we see in Fig.~\ref{fig2}
that the sensitivity in KASKA
changes only to $\sin^2{2\theta}\simeq 0.03$
even for $\Delta m^2 = 2\times 10^{-3}\eV^2$.

\begin{figure}
\hglue -1.8cm
\includegraphics[scale=0.6]{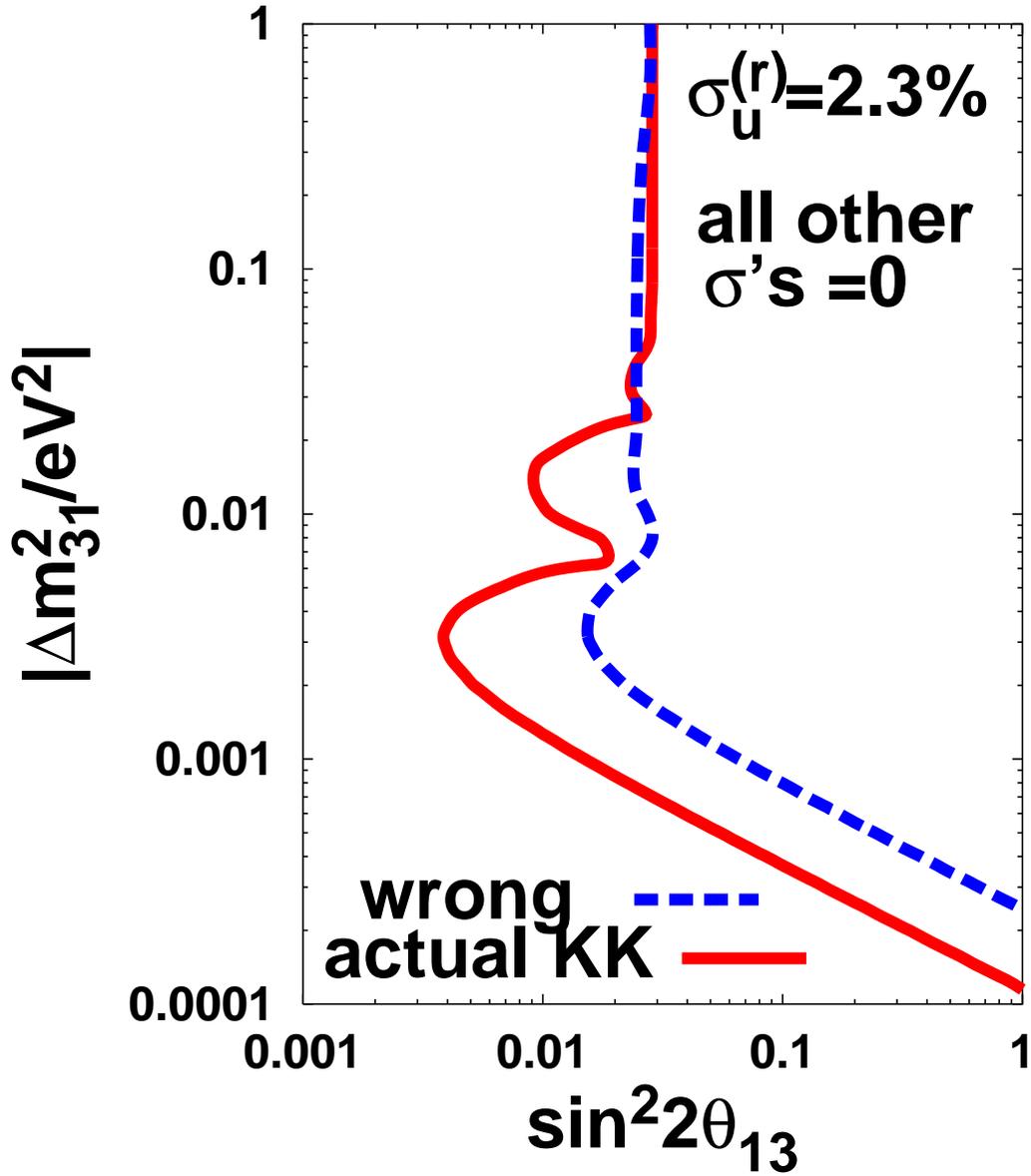}
\caption{
The comparison of the locations of the near detectors
in the KASKA plan.  The solid (dashed) line is for the
actual (hypothetical) KASKA plan with the location
depicted in Fig.\ref{fig1}(c) (Fig.\ref{fig1}(f)), respectively.
}
\label{fig3}
\end{figure}

Now let us see the effects of multiple sources
on the systematic limit 
$\left(\sin^22\theta\right)_{\text{limit}}^{\text{sys~only}}$
in the KASKA plan.
 Since the effect comes with $\sigma_{\text{u}}^{(\text{r})}$
(compare (\ref{chi4}) with (\ref{chi0})),
we can ignore other errors.
 The systematic limit for the case
is presented in Fig.~\ref{fig3} with solid line;
 $\sigma_{\text{u}}^{(\text{r})}$ is assumed to be 2.3\%.
 Note that
the systematic limit is extremely close to zero
for one reactor case because of ideal near-far cancellation.
 The solid line in Fig.~\ref{fig3}
shows, however, that about 0.4\% error remains.
 Thus,
we find that the effect of the nature of multiple sources
is only about 0.4\%.
 It is vary small compared to the relative normalization error
$\sqrt{2} \sigma_{\text{u}} (\simeq 0.8\%)$ of the detectors.%
\footnote{
Note that different types of errors are combined
as the sum of squared: $0.8^2 + 0.4^2 \simeq 0.89^2$.
}
 Once we know that two near detectors are sufficient
for near-far cancellation with the Kashiwazaki-Kariwa
nuclear power station,
we should investigate the optimal locations of the near detectors
for the cancellation.
 While the two near detectors are assumed to
be very close to the reactors
in the ideal limit (Fig.~\ref{fig1}(d)), the near detectors in the actual
KASKA case cannot be too close to the reactors.
 If each of the two near detectors is too close to a reactor,
it is impossible to cancel
the uncorrelated error of the flux from other reactors,
namely $\sigma_{\text{u}}^{\text(r)}$.
 Hence,
the optimization of the locations of two near detectors
at the Kashiwazaki-Kariwa site is nontrivial.
 At first,
in order to see the importance of the location of near detectors,
we compare the systematic limits
of the actual KASKA plan (Fig.~\ref{fig1}(c))
with that of a hypothetical experiment (Fig.~\ref{fig1}(f))
in Fig.~\ref{fig3};
 In the hypothetical experiment,
one near detector is very close to the reactor \#1 in the
first cluster while the other detector is very close to the reactor \#5
in the second one.
 In Fig.~\ref{fig3}
we use $\sigma_{\text{u}}^{\text{(r)}}=2.3\%$ and all other
systematic errors are set to zero
because $\sigma_{\text{u}}^{\text{(r)}}$
dominates the difference of near-far cancellations
in the two cases.
For $\Delta m^2=2.5\times10^{-3}\eV^2$, we obtain
$\left(\sin^22\theta\right)_{\text{limit}}^{\text{sys~only}}\simeq
4\times10^{-3}~(2\times10^{-2})$ for the actual KASKA plan
(the hypothetical case).
 In the hypothetical case, the sensitivity is
deteriorated because the correlated error
$\sigma_{\text{u}}^{\text{(r)}}$ is not canceled sufficiently
and about 2\% error remains.

\begin{figure}
\hglue -0.5cm
\includegraphics[scale=0.68,angle=-90]{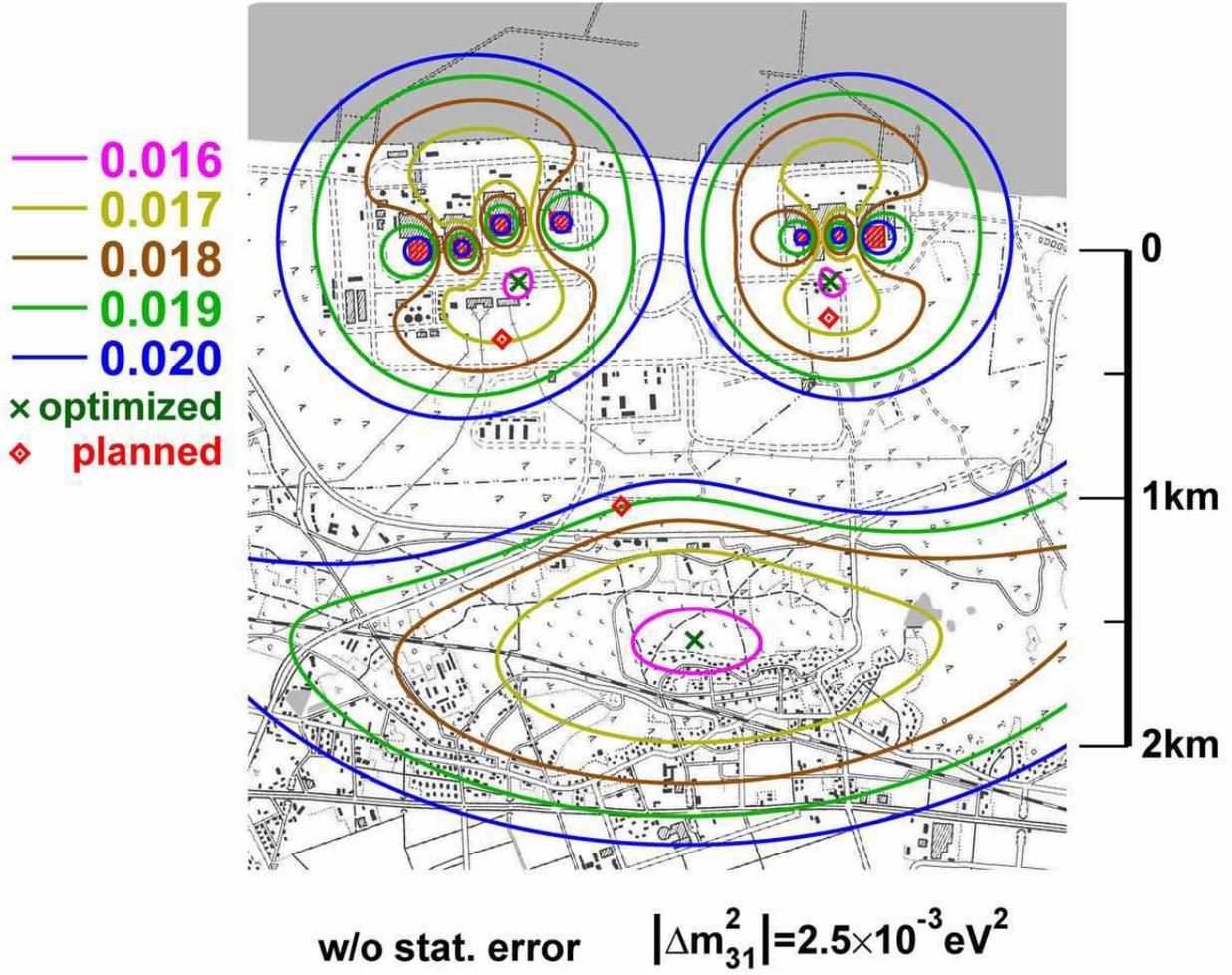}
\caption{
The contour plots of the systematic limit
on $\sin^22\theta_{13}$
in the KASKA experiment.
The optimized and currently planned positions of the detectors
are also depicted.
When the contour for each detector is plotted,
it is assumed that other detectors are located in
the optimized positions.
}
\label{figkka}
\end{figure}

\begin{figure}
\hglue -0.5cm
\includegraphics[scale=0.68,angle=-90]{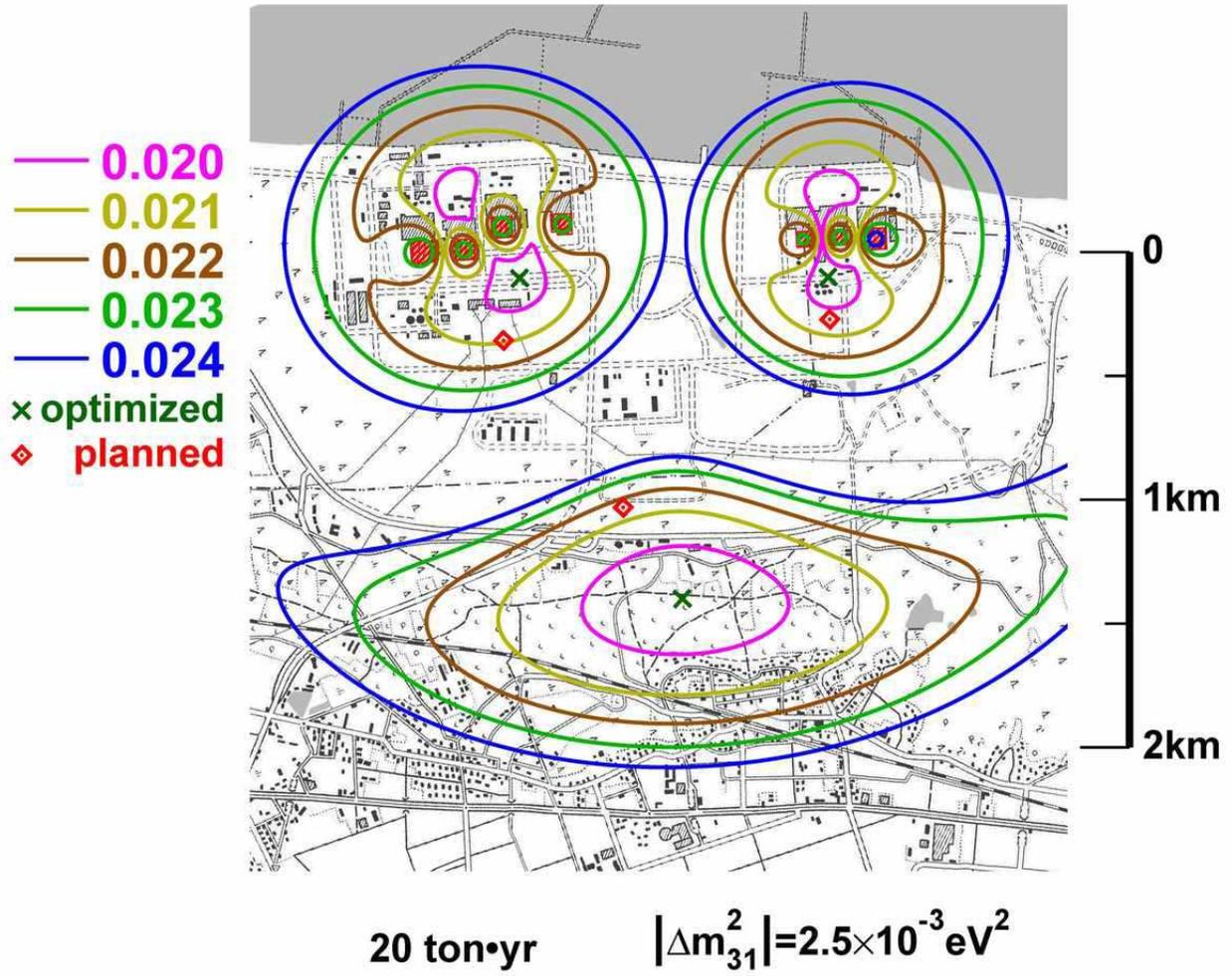}
\caption{
The same contour plot of the sensitivity to $\sin^22\theta_{13}$
as Fig.~\ref{figkka} with the data size of 20ton$\cdot$yr.
}
\label{figkkb}
\end{figure}

We can extend this analysis to that with the arbitrary position of each
detector.  To do the analysis, we first obtain the optimized
positions of the detectors.  And then we examine the sensitivity
to $\sin^22\theta$ by varying the position of each detector,
leaving the locations of the remaining detectors in the optimized ones.
The results are given by the contour plots
in Fig.~\ref{figkka} without statistical errors
and in Fig.~\ref{figkkb} with the data size of 20 ton$\cdot$yr,
where the reference values in Eqs. (\ref{error1}) and (\ref{error2})
are used.
In these figures the locations of the detectors are
also depicted for the optimized case and for the
currently planned case.  From these two figures we observe that
the distance between each near detector and the reactors in
each cluster is approximately (300$\pm$130)m
in the optimized case.  This results
in slightly poorer sensitivity to $\sin^22\theta$,
compared with the hypothetical single reactor case with a
near detector which is arbitrarily close to the reactor
(See Eq.~(\ref{sens1}) with $L_{\text{n}}=0$).
From Figs.~\ref{fig3},~\ref{figkka},~\ref{figkkb},
we see that the positions of the near detectors in the KASKA
plan are appropriate for the near-far cancellation and almost optimized.
 Therefore, it does not suffer from the multi-reactor nature
such as the variety of reactor powers
or distances between the reactors and the near detectors.

\section{Discussion and Conclusion}
Using the analytical method, we estimated
the systematic limits (sensitivity without statistical error)
on the neutrino oscillation parameter
$\sin^22\theta_{13}$ in various setups of reactor experiments.
In the simplest case, where there is one reactor and two
detectors, the correlated systematic error is canceled.
 In the case of multiple $n_r$ reactors,
we showed that the correlated systematic errors
of the reactors and the detectors are canceled as naive expectation
if the number of detectors $n_d$ is sufficient ($n_d = n_r+1$).
 We found that multiple reactors and detectors set up
has an advantage of the reduction of the remaining uncorrelated error
if the set up is appropriate for the near-far cancellation
of correlated errors.
 On the other hand,
we explicitly showed
that the contribution to the sensitivity to $\sin^22\theta_{13}$
from the the uncorrelated error
of the flux, which controls the multi-reactor nature, is negligibly small
although there are only three detectors for seven reactors ($n_d < n_r+1$). 
 The only disadvantage of experiments with $n_d<n_r+1$
is that one cannot put the near detectors
arbitrarily close to one of the reactors (even if one neglects
the technical difficulties), because that would ruin the
cancellation of the uncorrelated error of the flux,
as we have seen explicitly
in the KASKA case.
 We presented also the optimal positions of detectors
in the KASKA plan;
 The planned position of near detectors are close to the optimal ones
although it is better if the baseline length for the far detector
becomes longer beyond the bound of the power station site.
In all cases studies here, it is the factor
$\sigma_{\text{u}}/\left[\left\langle\sin^2\left(
\Delta m^2L_{\text{f}}/4E
\right)\right\rangle-\left\langle\sin^2\left(
\Delta m^2L_{\text{n}}/4E\right)\right\rangle\right]$
that sets the systematic limit
on $\sin^22\theta_{13}$, and hence it is quite important to estimate
$\sigma_{\text{u}}$ carefully.
 The factor
$\left(\sin^22\theta\right)_{\text{limit}}^{\text{sys~only}}/\sigma_{\text{u}}$
seems to be a good measure of the power of a reactor experiment
almost independently of assumptions of error sizes.

\appendix

\section{Derivation of the covariance matrix\label{appendix1}}
In this appendix we first show that the form of $\chi^2$ which is
expressed as the minimum of the function of the $\alpha$ variables
with respect to these variables leads to the
form of $\chi^2$ which is bilinear in the ratio $m_j/t_j$ of the
measured value $m_j$ divided by the theoretical prediction $t_j$.
This has been known in the literature
\cite{Stump:2001gu,Botje:2001fx,Pumplin:2002vw,Fogli:2002pt}
as the equivalence between the so-called pull approach and the
covariance matrix approach.  And then we show
that the same job can be done by integration of $\exp(-\chi^2/2)$
over the variables $m_j/t_j$.

In the cases which we are considering, the correlated systematic
errors $t_1^2\sigma_{\text{u}\,1}^2, \cdots, t_n^2\sigma_{\text{u}\,\ell}^2$
are introduced by the variables
\begin{eqnarray}
\vec{\alpha}\equiv\left(\begin{array}{cc}
\alpha_1 \\
\vdots\\
\alpha_\ell
\end{array}
\right).
\nonumber
\end{eqnarray}
Introducing the notation
\begin{eqnarray}
y_j&\equiv& \displaystyle{m_j \over t_j}-1,\nonumber\\
\vec{y}&\equiv&\left(\begin{array}{cc}
y_1 \\
\vdots\\
y_n
\end{array}
\right),
\nonumber
\end{eqnarray}
$\chi^2$ can be written as
\begin{eqnarray}
\chi^2&=&\min_{\vec{\alpha}}
\left[\left(\vec{y}-H\vec{\alpha}\right)^T D_{\text{u}}^{-1}
\left(\vec{y}-H\vec{\alpha}\right)
+\vec{\alpha}^T D_{\text{c}}^{-1}\vec{\alpha}
\right]\nonumber\\
&=&\min_{\vec{\alpha}}\left[
\left(\vec{\alpha}-A^{-1}HD_{\text{u}}^{-1}\vec{y}\right)^T
A\left(\vec{\alpha}-A^{-1}HD_{\text{u}}^{-1}\vec{y}\right)
+\vec{y}^T \left(D_{\text{u}}^{-1}-D_{\text{u}}^{-1}
HA^{-1}H^TD_{\text{u}}^{-1}\right)\vec{y}
\right]\nonumber\\
&=&\vec{y}^{\,T} \left(D_{\text{u}}^{-1}-D_{\text{u}}^{-1}
HA^{-1}H^TD_{\text{u}}^{-1}\right)\vec{y},
\label{pull1}
\end{eqnarray}
where
\begin{eqnarray}
H&\equiv&\left(\begin{array}{ccc}
1&\cdots\cdots&1\\
\vdots&&\vdots\\
1&\cdots\cdots&1
\end{array}\right),\nonumber
\end{eqnarray}
is an $n\times \ell$ matrix,
\begin{eqnarray}
D_{\text{u}}&\equiv&\mbox{\rm diag}\left(
\sigma_{\text{u}\,1}^2,\cdots,\sigma_{\text{u}\,n}^2
\right)\nonumber
\end{eqnarray}
is an $n\times n$ diagonal matrix whose element is
the normalized uncorrelated systematic error $\sigma_{\text{u}\,j}^2$
for the variable $m_j$ ($j=1,\cdots,n$),
\begin{eqnarray}
D_{\text{c}}&\equiv&\mbox{\rm diag}\left(
\sigma_{\text{c}\,1}^2,\cdots,\sigma_{\text{c}\,\ell}^2
\right)
\nonumber
\end{eqnarray}
is an $\ell\times \ell$ diagonal matrix whose element is
the normalized correlated systematic error  $\sigma_{\text{c}\,j}^2$
for the variable $\alpha_j$ ($j=1,\cdots,\ell$), and
we have defined
\begin{eqnarray}
A\equiv D_{\text{c}}^{-1}+H^TD_{\text{u}}^{-1}H.
\nonumber
\end{eqnarray}
Note that we can incorporate the effect of the statistical errors
in our formalism by redefining
$D_u\rightarrow \mbox{\rm diag}(\sigma_{\text{u}\,1}^2
+1/\sqrt{t_1},\cdots,\sigma_{\text{u}\,1}^2
+1/\sqrt{t_n})$, although we do not discuss the statistical errors
in the present paper.
From Eq.~(\ref{pull1}) we see that the covariance matrix $V$
is given by
\begin{eqnarray}
V=\left(D_{\text{u}}^{-1}-D_{\text{u}}^{-1}
HA^{-1}H^TD_{\text{u}}^{-1}\right)^{-1}.
\nonumber
\end{eqnarray}
We could prove by brute force that $V$ can be written as
\begin{eqnarray}
V=D_{\text{u}}+HD_{\text{c}}H^T,
\label{v}
\end{eqnarray}
but it is much easier to prove it by expressing the
matrix element $V_{ij}$ as the integral of $\exp(-\chi^2/2)$
over the variables $y_j$.

First of all, let us prove that the matrix element $V_{ij}$ can be
written as
\begin{eqnarray}
V_{ij}={\cal N}_y\int d\vec{y}\, y_i y_j\,
\exp\left(-{1 \over 2}
\vec{y}^{\,T}V^{-1}\vec{y}
\right),
\label{vij}
\end{eqnarray}
where ${\cal N}_y$ is the normalization constant defined by
\begin{eqnarray}
{\cal N}_y^{-1}\equiv
\int d\vec{y}\,
\exp\left(-{1 \over 2}
\vec{y}^{\,T}V^{-1}\vec{y}
\right).
\nonumber
\end{eqnarray}
Proof of Eq.~(\ref{vij}) goes as follows.
Diagonalizing the covariance matrix $V$, which is real symmetric,
by an orthogonal matrix ${\cal O}$
\begin{eqnarray}
V={\cal O}^TD{\cal O}\equiv{\cal O}^T\mbox{\rm diag}\left(
v_1,\cdots,v_n\right){\cal O},
\nonumber
\end{eqnarray}
the exponent can be rewritten as
\begin{eqnarray}
\vec{y}^{\,T}V^{-1}\vec{y}=\vec{y}^{\,T}{\cal O}^TD^{-1}{\cal O}\vec{y}
\equiv \vec{y'}^{\,T}D^{-1}\vec{y'},
\nonumber
\end{eqnarray}
so that we have
\begin{eqnarray}
&{\ }&{\cal N}_y\int d\vec{y}\, y_i y_j\,
\exp\left(-{1 \over 2}
\vec{y}^{\,T}V^{-1}\vec{y}
\right)\nonumber\\
&=&{\cal N}_y\int d\vec{y'}\, \left({\cal O}^T\vec{y'}\right)_i 
\left({\cal O}^T\vec{y'}\right)_j\,
\exp\left(-{1 \over 2}
\vec{y'}^{\,T}D^{-1}\vec{y'}
\right)\nonumber\\
&=&{\cal N}_y\int d\vec{y'}\, \left({\cal O}\right)_{ki}y'_k
\left({\cal O}\right)_{l j}y'_l\,
\exp\left(-{1 \over 2}
\sum_{i=1}^n\displaystyle{y_j^{'2} \over v_j}
\right)\nonumber\\
&=&\left({\cal O}\right)_{ki}\left(D\right)_{kl}
\left({\cal O}\right)_{lj}\nonumber\\
&=&\left({\cal O}^TD{\cal O}\right)_{ij}\nonumber\\
&=&V_{ij}.
\nonumber
\end{eqnarray}
Thus Eq.~(\ref{vij}) is proved.

Now Eq.~(\ref{vij}) can be simplified by expressing as the
integral over the variables $\vec{\alpha}$ of the
original $\chi^2$:
\begin{eqnarray}
V_{ij}&=&{\cal N}_y\int d\vec{y}\, y_i y_j\,
\exp\left(-{1 \over 2}
\vec{y}^{\,T}V^{-1}\vec{y}
\right)\nonumber\\
&=&{\cal N}_y{\cal N}_\alpha\int d\vec{y}\,\int d\vec{\alpha}\, y_i y_j\,
\exp\left\{-{1 \over 2}\left[
\left(\vec{\alpha}-A^{-1}HD_{\text{u}}^{-1}\vec{y}\right)^T
A\left(\vec{\alpha}-A^{-1}HD_{\text{u}}^{-1}\vec{y}\right)
\right.\right.\nonumber\\
&{\ }&\left.\left.+\vec{y}^T \left(D_{\text{u}}^{-1}-D_{\text{u}}^{-1}
HA^{-1}H^TD_{\text{u}}^{-1}\right)\vec{y}
\right]\right\}\nonumber\\
&=&{\cal N}_y{\cal N}_\alpha\int d\vec{y}\,\int d\vec{\alpha}\, y_i y_j\,
\exp\left\{-{1 \over 2}
\left[\left(\vec{y}-H\vec{\alpha}\right)^T D_{\text{u}}^{-1}
\left(\vec{y}-H\vec{\alpha}\right)
+\vec{\alpha}^T D_{\text{c}}^{-1}\vec{\alpha}
\right]\right\},
\label{vij2}
\end{eqnarray}
where we have used Eq.~(\ref{pull1}),
the normalization constant ${\cal N}_\alpha$ is defined by
\begin{eqnarray}
{\cal N}_\alpha^{-1}&\equiv&
\int d\vec{\alpha}\,
\exp\left\{-{1 \over 2}
\left[
\left(\vec{\alpha}-A^{-1}HD_{\text{u}}^{-1}\vec{y}\right)^T
A\left(\vec{\alpha}-A^{-1}HD_{\text{u}}^{-1}\vec{y}\right)
\right]\right\},
\nonumber
\end{eqnarray}
and ${\cal N}_y$ and ${\cal N}_\alpha$ are related by
\begin{eqnarray}
{\cal N}_y{\cal N}_\alpha=\left(2\pi\right)^{(n+\ell)/2}
\left(\det D_{\text{u}}\right)^{1/2}\left(\det D_{\text{c}}\right)^{1/2}.
\nonumber
\end{eqnarray}
Eq.~(\ref{vij2}) can be easily calculated by shifting the
variable $\vec{y}\rightarrow\vec{y''}\equiv\vec{y}-H\vec{\alpha}$:
\begin{eqnarray}
V_{ij}&=&{\cal N}_y{\cal N}_\alpha\int d\vec{y''}\,\int d\vec{\alpha}\,
\left(\vec{y''}-H\vec{\alpha}\right)_i 
\left(\vec{y''}-H\vec{\alpha}\right)_j\nonumber\\
&{\ }&\exp\left\{-{1 \over 2}
\left[\left(\vec{y}-H\vec{\alpha}\right)^T D_{\text{u}}^{-1}
\left(\vec{y}-H\vec{\alpha}\right)
+\vec{\alpha}^T D_{\text{c}}^{-1}\vec{\alpha}
\right]\right\}\nonumber\\
&=&\left(D_{\text{u}}\right)_{ij}
+\left(H\right)_{ik}\left(H\right)_{jl}\left(D_{\text{c}}\right)_{kl}\nonumber\\
&=&\left(D_{\text{u}}+HD_{\text{c}}H^T\right)_{ij}.
\nonumber
\end{eqnarray}
Hence Eq.~(\ref{v}) is proved.

In the case of one reactor with one detector (cf.\ Eq.~(\ref{chi0})),
we have
\begin{eqnarray}
D_{\text{u}}&=&\sigma_{\text{u}}^2\nonumber\\
D_{\text{c}}&=&{\rm diag}\left(\sigma_{\text{c}}^2,
(\sigma_{\text{c}}^{\text{(r)}})^2,
(\sigma_{\text{u}}^{\text{(r)}})^2\right)\nonumber\\
H&=&(1,1,1),
\nonumber
\end{eqnarray}
so the covariance matrix is
\begin{eqnarray}
V=D_{\text{u}}+HD_{\text{c}}H^T=\sigma_{\text{u}}^2+
\sigma_{\text{c}}^2+(\sigma_{\text{c}}^{\text{(r)}})^2
+(\sigma_{\text{u}}^{\text{(r)}})^2.
\nonumber
\end{eqnarray}
In the case of one reactor with two detectors (cf.\ Eq.~(\ref{1r2d})),
we have
\begin{eqnarray}
D_{\text{u}}&=&\sigma_{\text{u}}^2\,\left(\begin{array}{cc}
1&0\\
0&1\\
\end{array}
\right)\nonumber\\
D_{\text{c}}&=&{\rm diag}\left(\sigma_{\text{c}}^2,
(\sigma_{\text{c}}^{\text{(r)}})^2,
(\sigma_{\text{u}}^{\text{(r)}})^2\right)\nonumber\\
H&=&\left(\begin{array}{ccc}
1&1&1\\
1&1&1\\
\end{array}
\right),
\nonumber
\end{eqnarray}
so that we obtain
\begin{eqnarray}
\hspace*{-10mm}
V=D_{\text{u}}+HD_{\text{c}}H^T=
\left(
\displaystyle
\begin{array}{rr}
\sigma^2_{\text{u}}+\sigma^2_{\text{c}}
+(\sigma^{\text{(r)}}_{\text{u}})^2
+(\sigma^{\text{(r)}}_{\text{c}})^2\hspace*{5mm}
& \quad\sigma^2_{\text{c}}
+(\sigma^{\text{(r)}}_{\text{u}})^2
+(\sigma^{\text{(r)}}_{\text{c}})^2\\
\quad\sigma^2_{\text{c}}
+(\sigma^{\text{(r)}}_{\text{u}})^2
+(\sigma^{\text{(r)}}_{\text{c}})^2\hspace*{5mm} &
\sigma^2_{\text{u}}+\sigma^2_{\text{c}}
+(\sigma^{\text{(r)}}_{\text{u}})^2
+(\sigma^{\text{(r)}}_{\text{c}})^2
\end{array}
\right).
\nonumber
\end{eqnarray}
In the case of $n_r$ reactors with one detector (cf.\ Eq.~(\ref{chi4})),
we have
\begin{eqnarray}
D_{\text{u}}&=&\sigma_{\text{u}}^2\nonumber\\
D_{\text{c}}&=&{\rm diag}\left[\sigma_{\text{c}}^2,
(\sigma_{\text{c}}^{\text{(r)}})^2,
(\sigma_{\text{u}}^{\text{(r)}})^2,
\cdots,(\sigma_{\text{u}}^{\text{(r)}})^2
\right]\nonumber\\
H&=&\left(1,1,{t_1 \over t},\cdots,{t_{n_r} \over t}
\right),
\nonumber
\end{eqnarray}
so that we obtain
\begin{eqnarray}
V&=&D_{\text{u}}+HD_{\text{c}}H^T\nonumber\\
&=&\sigma_{\text{u}}^2
+\left(1,1,{t_1 \over t},\cdots,{t_{n_r} \over t}
\right){\rm diag}\left[\sigma_{\text{c}}^2,
(\sigma_{\text{c}}^{\text{(r)}})^2,
(\sigma_{\text{u}}^{\text{(r)}})^2,
\cdots,(\sigma_{\text{u}}^{\text{(r)}})^2
\right]\left(\begin{array}{c}
1\\
1\\
{t_1 / t}\\
\vdots\\
{t_{n_r} / t}
\end{array}
\right)\nonumber\\
&=&\sigma_{\text{u}}^2+\sigma_{\text{c}}^2
+(\sigma_{\text{c}}^{\text{(r)}})^2
+(\sigma_{\text{u}}^{\text{(r)}})^2
\sum_{a=1}^{n_r} \left({t_a \over t}\right)^2.
\nonumber
\end{eqnarray}
In the case of $n_r$ reactors with $n_d$ detector (cf.\ Eq.~(\ref{chi5})),
we have
\begin{eqnarray}
D_{\text{u}}&=&\sigma_{\text{u}}^2\,I_{n_d}\nonumber\\
D_{\text{c}}&=&{\rm diag}\left[\sigma_{\text{c}}^2,
(\sigma_{\text{c}}^{\text{(r)}})^2,
(\sigma_{\text{u}}^{\text{(r)}})^2,
\cdots,(\sigma_{\text{u}}^{\text{(r)}})^2
\right]\nonumber\\
H&=&\left(\begin{array}{ccccc}
1&1&{t_1 \over t}&\cdots&{t_{n_r} \over t}\\
\vdots&\vdots&\vdots&\vdots&\vdots\\
1&1&{t_1 \over t}&\cdots&{t_{n_r} \over t}\\
\end{array}
\right),
\nonumber
\end{eqnarray}
where $I_{n_d}$ is an $n_d\times n_d$ unit matrix,
so that we obtain
\begin{eqnarray}
V&=&\sigma_{\text{u}}^2I_{n_d}
+\left(\begin{array}{ccccc}
1&1&{t_1 \over t}&\cdots&{t_{n_r} \over t}\\
\vdots&\vdots&\vdots&\vdots&\vdots\\
1&1&{t_1 \over t}&\cdots&{t_{n_r} \over t}\\
\end{array}
\right)
{\rm diag}\left[\sigma_{\text{c}}^2,
(\sigma_{\text{c}}^{\text{(r)}})^2,
(\sigma_{\text{u}}^{\text{(r)}})^2,
\cdots,(\sigma_{\text{u}}^{\text{(r)}})^2
\right]\left(\begin{array}{ccc}
1&\cdots&1\\
1&\cdots&1\\
{t_1 / t}&\cdots&{t_1 / t}\\
\vdots&\vdots&\vdots\\
{t_{n_r} / t}&\cdots&{t_{n_r} / t}
\end{array}
\right)\nonumber\\
&=&\sigma_{\text{u}}^2I_{n_d}
+\left[\sigma_{\text{c}}^2
+(\sigma_{\text{c}}^{\text{(r)}})^2
+(\sigma_{\text{u}}^{\text{(r)}})^2
\sum_{a=1}^{n_r} \left({t_a \over t}\right)^2\right]
H_{n_d},
\nonumber
\end{eqnarray}
where $H_{n_d}$ is an $n_d\times n_d$ matrix whose
elements are all 1 (cf.\ Eq.~(\ref{h})).

\section{Derivation of $\chi^2$ (\ref{chinrnr1d})
in the case with $n_r$ reactors and $n_r+1$ detectors\label{appendix2}}
$V$ in Eq.~(\ref{vnrnr1d}) can be diagonalized as
\begin{eqnarray}
&{\ }&U^{-1}VU\nonumber\\
&=&\mbox{\rm diag}\left\{
(n_r+1)\left[\sigma_{\text{c}}^2
+\left(\sigma_{\text{c}}^{(r)}\right)^2
+{\sigma_{\text{u}}^2 \over n_r}
+{\left(\sigma_{\text{u}}^{(r)}\right)^2 \over n_r}
\right],\sigma_{\text{u}}^2
+\left(\sigma_{\text{u}}^{(r)}\right)^2,
\cdots,\sigma_{\text{u}}^2
+\left(\sigma_{\text{u}}^{(r)}\right)^2,
\sigma_{\text{u}}^2
\right\},\nonumber
\end{eqnarray}
where $U$ is a unitary matrix defined by
\begin{eqnarray}
\hspace*{-20mm}
U&=&\left(\vec{u}^{(1)},\cdots,\vec{u}^{(n)}
\right),\nonumber\\
\vec{u}^{(1)}&\equiv&{1 \over \sqrt{N_r+1}}
\left(\begin{array}{c}
1\\
\vdots\\
1\end{array}\right),\quad
\vec{u}^{(2)}\equiv{1 \over \sqrt{2}}
\left(\begin{array}{r}
1\\
-1\\
0\\
\vdots\\
0\end{array}\right),\nonumber\\
\vec{u}^{(3)}&\equiv&{1 \over \sqrt{6}}
\left(\begin{array}{r}
1\\
1\\
-2\\
0\\
\vdots\\
0\end{array}\right),\cdots,\quad
\vec{u}^{(n_r+1)}\equiv{1 \over \sqrt{n_r(n_r+1)}}
\left(\begin{array}{r}
1{\ }\,\\
\vdots{\ }~\\
1{\ }\,\\
-n_r\end{array}\right).
\label{u}
\end{eqnarray}
Introducing the notation
\begin{eqnarray}
\vec{y}\equiv\left(\begin{array}{c}
m_1/t_1-1\\
\vdots\\
m_{n_r}/t_{n_r}-1\\
m_{n_r+1}/t_{n_r+1}-1\\
\end{array}
\right),
\nonumber
\end{eqnarray}
$\chi^2$ can be written as
\begin{eqnarray}
\chi^2&=&{\left(\vec{y}\cdot\vec{u}^{(1)}\right)^2 \over
\sigma_{\text{u}}^2
+(n_r+1)\left[\sigma_{\text{c}}^2
+\left(\sigma_{\text{c}}^{(r)}\right)^2
+\left(\sigma_{\text{u}}^{(r)}\right)^2/n_r
\right]}
+{\displaystyle\sum_{j=2}^{n_r}
\left(\vec{y}\cdot\vec{u}^{(j)}\right)^2 \over
\sigma_{\text{u}}^2
+\left(\sigma_{\text{u}}^{(r)}\right)^2}
+{\left(\vec{y}\cdot\vec{u}^{(n_r+1)}\right)^2 \over
\sigma_{\text{u}}^2}.
\nonumber
\end{eqnarray}
From Eq.~(\ref{u}) we have
$\vec{y}\cdot\vec{u}_1=D(L_{\text{f}})/\sqrt{n_r+1}$,
$\vec{y}\cdot\vec{u}_j=0~(j=2,\cdots,n_r)$,
$\vec{y}\cdot\vec{u}_{n_r+1}=-\sqrt{n_r/(n_r+1)}D(L_{\text{f}})$,
and we finally get
\begin{eqnarray}
\chi^2&=&\left[D(L_{\text{f}})\right]^2\left\{
{n_r \over n_r+1}{1 \over \sigma_{\text{u}}^2}
+{1 \over n_r+1} {1\over
\sigma_{\text{u}}^2
+(n_r+1)\left[\sigma_{\text{c}}^2
+\left(\sigma_{\text{c}}^{(r)}\right)^2
+\left(\sigma_{\text{u}}^{(r)}\right)^2/n_r
\right]}
\right\}.
\nonumber
\end{eqnarray}

\section{Derivation of the systematic limit in the KASKA
plan\label{appendix3}}
It is easy to show that the diagonalized
matrix out of
the covariance matrix $V$ (\ref{v7r3d}) is
\begin{eqnarray}
\hspace*{-10mm}
\mbox{\rm diag}(\sigma_{\text{u}}^2,
\sigma_{\text{u}}^2+\Lambda_+,
\sigma_{\text{u}}^2+\Lambda_-),
\label{diag1}
\end{eqnarray}
where
\begin{eqnarray}
\hspace*{-10mm}
\Lambda_\pm&\equiv&{3 \over 2}
\left[\sigma_{\text{c}}^2+(\sigma^{\text{(r)}}_{\text{c}})^2\right]
+{61 \over 168}(\sigma^{\text{(r)}}_{\text{u}})^2\nonumber\\
&\pm&\displaystyle{1 \over 2}
\left\{9\left[\sigma_{\text{c}}^2
+(\sigma^{\text{(r)}}_{\text{c}})^2\right]^2
+{5 \over 6}\left[\sigma_{\text{c}}^2
+(\sigma^{\text{(r)}}_{\text{c}})^2\right]
(\sigma^{\text{(r)}}_{\text{u}})^2
+\left({13 \over 84}\right)^2
(\sigma^{\text{(r)}}_{\text{u}})^4\right\}^{1/2}.
\nonumber
\end{eqnarray}
The corresponding eigenvectors are
\begin{eqnarray}
\vec{u}^{(1)}={1 \over \sqrt{74}}
\left(
\begin{array}{c}
-4\\
-3\\
7\\
\end{array}
\right),\vec{u}^{(2)}={\cal N}_+
\left(
\begin{array}{c}
-\Lambda_++46\eta\\
-\Lambda_++53\eta\\
-\Lambda_++49\eta
\end{array}
\right),\vec{u}^{(3)}={\cal N}_-
\left(
\begin{array}{c}
-\Lambda_-+46\eta\\
-\Lambda_-+53\eta\\
-\Lambda_-+49\eta
\end{array}
\right),
\nonumber
\end{eqnarray}
where
\begin{eqnarray}
\eta\equiv{1 \over 84}\,
{(\sigma_{\text{u}}^{\text{(r)}})^2 \over
\sigma_{\text{c}}^2+(\sigma_{\text{c}}^{\text{(r)}})^2},
\nonumber
\end{eqnarray}
and ${\cal N}_\pm$ are the normalization constants.
Hence we get
\begin{eqnarray}
\chi^2&=&\sin^42\theta
\left\{\displaystyle{49 \over 74}
\displaystyle{\left[D(L_{\text{f}})\right]^2 \over \sigma_{\text{u}}^2}
+\displaystyle{\left[u_3^{(2)}D(L_{\text{f}})\right]^2 \over 
\sigma_{\text{u}}^2+\Lambda_-}
+\displaystyle{\left[u_3^{(3)}D(L_{\text{f}})\right]^2 \over 
\sigma_{\text{u}}^2+\Lambda_+}
\right\}\nonumber\\
&\simeq&\sin^42\theta\,
\displaystyle{49 \over 74}\,
\displaystyle{\left[D(L_{\text{f}})\right]^2 \over 
\sigma_{\text{u}}^2}.
\nonumber
\end{eqnarray}

In the actual KASKA case, using the
reference values (\ref{error1}) and (\ref{error2}),
from numerical calculations
we obtain the diagonalized covariance matrix
\begin{eqnarray}
\mbox{\rm diag}(\lambda_1,
\lambda_2,
\lambda_3)=\sigma_{\text{u}}^2\mbox{\rm diag}
(1.044,4.651,90.878),
\nonumber
\end{eqnarray}
and the corresponding three eigenvectors
\begin{eqnarray}
\vec{u}^{(1)}\equiv
\left(
\begin{array}{c}
u_1^{(1)}\\
u_2^{(1)}\\
u_3^{(1)}\\
\end{array}
\right)=\left(
\begin{array}{c}
-0.4975\\
-0.3114\\
0.8097\\
\end{array}
\right),~
\vec{u}^{(2)}=\left(
\begin{array}{c}
0.6503\\
-0.7516\\
0.1105\\
\end{array}
\right),~
\vec{u}^{(3)}=\left(
\begin{array}{c}
0.5741\\
0.5815\\
0.5764\\
\end{array}
\right).
\nonumber
\end{eqnarray}
$\chi^2$ can be written as
\begin{eqnarray}
\chi^2=
{\left(\vec{y}\cdot\vec{u}^{(1)}\right)^2 \over \lambda_1}
+{\left(\vec{y}\cdot\vec{u}^{(2)}\right)^2 \over \lambda_2}
+{\left(\vec{y}\cdot\vec{u}^{(3)}\right)^2 \over \lambda_3}
\simeq
{\left(\vec{y}\cdot\vec{u}^{(1)}\right)^2 \over \lambda_1},
\nonumber
\end{eqnarray}
where $\vec{y}$ in this case is given by
\begin{eqnarray}
\vec{y}\equiv\left(\begin{array}{c}
{m_1 \over t_1}-1\\
{m_2 \over t_2}-1\\
{m_3 \over t_3}-1\\
\end{array}
\right)=-\sin^22\theta\left(\begin{array}{c}
\displaystyle\sum_{a=1}^7{t_{a1} \over t_1}
D(L_{a1})\\
\displaystyle\sum_{a=1}^7{t_{a2} \over t_2}
D(L_{a2})\\
\displaystyle\sum_{a=1}^7{t_{a3} \over t_3}
D(L_{a3})\\
\end{array}
\right)\equiv
-\sin^22\theta\left(\begin{array}{c}
\overline{D(L_{1})}\\
\overline{D(L_{2})}\\
\overline{D(L_{3})}\\
\end{array}
\right)
\nonumber
\end{eqnarray}
using the definition of $D(L)$ in Eq.~(\ref{d}) and
$\overline{D(L_j)}$ in Eq.~(\ref{overlined}).
Hence we get
\begin{eqnarray}
\chi^2\simeq{\sin^42\theta \over 1.04\sigma_{\text{u}}^2}
\left[\overline{D(L_3)}u^{(1)}_3+
\overline{D(L_1)}u^{(1)}_1+
\overline{D(L_2)}u^{(1)}_2
\right]^2,
\nonumber
\end{eqnarray}
from which Eq.~(\ref{limitkk}) follows.

%%%%%%%%%%%%%%%% acknowledgments %%%%%%%%%%%%%%%%
\begin{acknowledgments}
This work was supported in part by Grants-in-Aid for Scientific Research
No.\ 11204015, No.\ 16540260 and No.\ 16340078, Japan Ministry
of Education, Culture, Sports, Science, and Technology,
and the Research Fellowship
of Japan Society for the Promotion of Science
for young scientists.
\end{acknowledgments}

%%%%%%%%%%%%%%%%%%%%%%%%%% Bibliography %%%%%%%%%%%%%%%%%%%%%%%%%%%%%%

\newpage
\hglue -1.8cm
\begin{table}
\vglue 0.cm
\hglue -1.8cm
\begin{tabular}{|r|r|r|r|r|}
\hline
Reactors($a$) & Power/GWth & $L_{a\,1}$/m& $L_{a\,2}$/m
& $L_{a\,3}$/m\\
\hline
1&3.293&482 & 1663 & 1309 \\
2&3.293&401 & 1504 & 1224 \\
3&3.293&458 & 1374 & 1233 \\
4&3.293&524 & 1149 & 1169 \\
5&3.293&1552 &  371 & 1484 \\
6&3.926&1419 &  333 & 1397 \\
7&3.926&1280 &  340 & 1306 \\
\hline
\end{tabular}
\caption{The Power of the the reactors in GW$_{\text{th}}$
and the distance $L_{aj}$ in meters from the three detectors
($j$=1(near), 2(near), 3(far))
to each reactor ($a=1,\cdots,7$).
The powers of reactors are listed in  
http://www.tepco.co.jp/kk-np/index-j.html.
The positions of the reactors were read from a map and are subject to
a few meters of inaccuracy.  However, such inaccuracy hardly affects
the estimation of the systematic errors in the text.}
\label{table1}
\end{table}

\hglue -0.2cm
\begin{table}
\vglue 2cm
\begin{tabular}{|r|r|r|r|}
\hline
Reactors($a$)&$t_{a1}/t_1$&$t_{a2}/t_2$&$t_{a3}/t_3$\\
\hline
 1 & 0.208 & 0.012 & 0.133 \\
 2 & 0.301 & 0.015 & 0.152 \\
 3 & 0.231 & 0.017 & 0.149 \\
 4 & 0.176 & 0.025 & 0.166 \\
 5 & 0.020 & 0.239 & 0.103 \\
 6 & 0.029 & 0.353 & 0.139 \\
 7 & 0.035 & 0.339 & 0.159 \\
\hline
\end{tabular}
\vglue 0.cm
\label{tab:error}
\caption{The fraction $t_{aj}/t_j$ of the yields at each
detector $j$ from the reactor $a$ in the KASKA plan.}
\label{table2}
\end{table}

\newpage

\newpage
\newpage
\newpage
\newpage

\end{document}